\documentclass[12pt,a4paper]{article}

\usepackage{ifthen} 
\newboolean{pdflatex}
\setboolean{pdflatex}{true} 

\newboolean{articletitles}
\setboolean{articletitles}{true} 

\newboolean{uprightparticles}
\setboolean{uprightparticles}{false} 
\usepackage{overpic}
\usepackage{multirow}

\def\paperauthors{LHCb collaboration} 
\def\paperasciititle{Observation of new excited Sigmac0 states in B- to Lambdac+ pbar pi- decay} 
\def\papertitle{Observation of new excited $\mathchar"7106_{c}^{0}$ states in the  ${\ensuremath{\B^-}}\xspace\rightarrow {\ensuremath{\mathchar"7103^+_\ensuremath{c}}}\xspace\overline{p}\pi^{-}$ decay} 
\def\paperkeywords{{High Energy Physics}, {LHCb}} 
\def\papercopyright{\the\year\ CERN for the benefit of the LHCb collaboration} 
\def\paperlicence{CC BY 4.0 licence}
\def\paperlicenceurl{https://creativecommons.org/licenses/by/4.0/}

\newif\ifEnableSectionTOCLinks
\EnableSectionTOCLinksfalse 

\usepackage[top=1in, bottom=1.25in, left=1in, right=1in]{geometry}

\usepackage{microtype}
\usepackage{lineno}  
\usepackage{xspace} 
\usepackage{caption} 

\usepackage{graphicx}  
\usepackage{color}
\usepackage{colortbl}
\graphicspath{{./figs/}} 

\usepackage{amsmath} 
\usepackage{amssymb}
\usepackage{amsfonts}
\usepackage{upgreek} 

\newcommand*\patchAmsMathEnvironmentForLineno[1]{%
\expandafter\let\csname old#1\expandafter\endcsname\csname #1\endcsname
\expandafter\let\csname oldend#1\expandafter\endcsname\csname
end#1\endcsname
 \renewenvironment{#1}%
   {\linenomath\csname old#1\endcsname}%
   {\csname oldend#1\endcsname\endlinenomath}%
}
\newcommand*\patchBothAmsMathEnvironmentsForLineno[1]{%
  \patchAmsMathEnvironmentForLineno{#1}%
  \patchAmsMathEnvironmentForLineno{#1*}%
}
\AtBeginDocument{%
\patchBothAmsMathEnvironmentsForLineno{equation}%
\patchBothAmsMathEnvironmentsForLineno{align}%
\patchBothAmsMathEnvironmentsForLineno{flalign}%
\patchBothAmsMathEnvironmentsForLineno{alignat}%
\patchBothAmsMathEnvironmentsForLineno{gather}%
\patchBothAmsMathEnvironmentsForLineno{multline}%
\patchBothAmsMathEnvironmentsForLineno{eqnarray}%
}

\usepackage[pdftex,
            pdfauthor={\paperauthors},
            pdftitle={\paperasciititle},
            pdfkeywords={\paperkeywords}]{hyperref}
\usepackage{hyperxmp}
\hypersetup{
    pdfcopyright={Copyright (C) \papercopyright},
    pdflicenseurl={\paperlicenceurl}
}

\usepackage[colorinlistoftodos,textsize=scriptsize]{todonotes}

\usepackage[bottom,flushmargin,hang,multiple]{footmisc}

\usepackage[all]{hypcap} 

\usepackage{xspace} 
\usepackage{upgreek}

\def\lhcb   {\mbox{LHCb}\xspace}

\def\babar  {\mbox{BaBar}\xspace}
\def\belle  {\mbox{Belle}\xspace}

\def\MagUp {\mbox{\em Mag\kern -0.05em Up}\xspace}

\ifthenelse{\boolean{uprightparticles}}%
{

 \def\Ppi         {\ensuremath{\uppi}\xspace}

 \def\PDelta      {\ensuremath{\Delta}\xspace}                 
 \def\PXi         {\ensuremath{\Xi}\xspace}                 
 \def\PLambda     {\ensuremath{\Lambda}\xspace}                 
 \def\PSigma      {\ensuremath{\Sigma}\xspace}                 
 \def\POmega      {\ensuremath{\Omega}\xspace}                 
 \def\PUpsilon    {\ensuremath{\Upsilon}\xspace}
 \let\oldPi\Pi
 \def\PPi         {\ensuremath{\oldPi}\xspace}

 \def\PB      {\ensuremath{\mathrm{B}}\xspace}                 
 \def\PD      {\ensuremath{\mathrm{D}}\xspace}                 

 \def\PK      {\ensuremath{\mathrm{K}}\xspace}                 

 \def\Pc      {\ensuremath{\mathrm{c}}\xspace}

 \def\Pp      {\ensuremath{\mathrm{p}}\xspace}                 

 \def\Ps      {\ensuremath{\mathrm{s}}\xspace}

 \def\thebaroffset{0.0em}
}
{

 \def\Ppi         {\ensuremath{\pi}\xspace}

 \mathchardef\PDelta="7101
 \mathchardef\PXi="7104
 \mathchardef\PLambda="7103
 \mathchardef\PSigma="7106
 \mathchardef\POmega="710A
 \mathchardef\PUpsilon="7107
 \mathchardef\PPi="7105
 \def\PB      {\ensuremath{B}\xspace}                 
 \def\PD      {\ensuremath{D}\xspace}                 

 \def\PK      {\ensuremath{K}\xspace}                 

 \def\Pc      {\ensuremath{c}\xspace}

 \def\Pp      {\ensuremath{p}\xspace}                 

 \def\Ps      {\ensuremath{s}\xspace}

 \def\thebaroffset{0.18em}
}
\newcommand{\offsetoverline}[2][\thebaroffset]{\kern #1\overline{\kern -#1 #2}}%

\makeatletter
\ifcase \@ptsize \relax
  \newcommand{\miniscule}{\@setfontsize\miniscule{4}{5}}
\or
  \newcommand{\miniscule}{\@setfontsize\miniscule{5}{6}}
\or
  \newcommand{\miniscule}{\@setfontsize\miniscule{5}{6}}
\fi
\makeatother

\DeclareRobustCommand{\optbar}[1]{\shortstack{{\miniscule (\rule[.5ex]{1.25em}{.18mm})}
  \\ [-.7ex] $#1$}}

\def\squark    {{\ensuremath{\Ps}}\xspace}

\def\cquark    {{\ensuremath{\Pc}}\xspace}

\def\pion   {{\ensuremath{\Ppi}}\xspace}

\def\pip    {{\ensuremath{\pion^+}}\xspace}
\def\pim    {{\ensuremath{\pion^-}}\xspace}

\def\kaon    {{\ensuremath{\PK}}\xspace}

\def\KorKbar {\kern \thebaroffset\optbar{\kern -\thebaroffset \PK}{}\xspace}

\def\Km      {{\ensuremath{\kaon^-}}\xspace}

\def\D       {{\ensuremath{\PD}}\xspace}

\def\DorDbar {\kern \thebaroffset\optbar{\kern -\thebaroffset \PD}\xspace}

\def\Dp      {{\ensuremath{\D^+}}\xspace}
\def\Dm      {{\ensuremath{\D^-}}\xspace}

\def\DpDm    {\ensuremath{\Dp {\kern -0.16em \Dm}}\xspace}

\def\B       {{\ensuremath{\PB}}\xspace}

\def\BorBbar {\kern \thebaroffset\optbar{\kern -\thebaroffset \PB}\xspace}

\def\Bd      {{\ensuremath{\B^0}}\xspace}

\def\BdorBdbar {\kern \thebaroffset\optbar{\kern -\thebaroffset \Bd}\xspace}

\def\Bub     {{\ensuremath{\B^-}}\xspace}

\def\Bm      {{\ensuremath{\Bub}}\xspace}

\def\Bs      {{\ensuremath{\B^0_\squark}}\xspace}

\def\BsorBsbar {\kern \thebaroffset\optbar{\kern -\thebaroffset \Bs}\xspace}

\def\Y#1S{\ensuremath{\PUpsilon{(#1S)}}\xspace}

\def\proton      {{\ensuremath{\Pp}}\xspace}
\def\antiproton  {{\ensuremath{\overline \proton}}\xspace}

\def\Deltares    {{\ensuremath{\PDelta}}\xspace}
\def\Deltaresbar {{\ensuremath{\overline \Deltares}}\xspace}
\def\Lz          {{\ensuremath{\PLambda}}\xspace}

\def\LorLbar     {\kern \thebaroffset\optbar{\kern -\thebaroffset \PLambda}\xspace}

\def\Sigmares    {{\ensuremath{\PSigma}}\xspace}

\def\Lc          {{\ensuremath{\Lz^+_\cquark}}\xspace}

\newcommand{\decay}[2]{\ensuremath{\mathinner{#1\!\to #2}}\xspace}

\def\to                 {\ensuremath{\rightarrow}\xspace}

\def\AT#1     {\ensuremath{A_{\mathrm{T}}^{#1}}\xspace}           

\def\C#1      {\ensuremath{\mathcal{C}_{#1}}\xspace}                       
\def\Cp#1     {\ensuremath{\mathcal{C}_{#1}^{'}}\xspace}                    
\def\Ceff#1   {\ensuremath{\mathcal{C}_{#1}^{\mathrm{(eff)}}}\xspace}        
\def\Cpeff#1  {\ensuremath{\mathcal{C}_{#1}^{'\mathrm{(eff)}}}\xspace}       
\def\Ope#1    {\ensuremath{\mathcal{O}_{#1}}\xspace}                       
\def\Opep#1   {\ensuremath{\mathcal{O}_{#1}^{'}}\xspace}                    

\newcommand{\aunit}[1]{\ensuremath{\text{\,#1}}}       

\newcommand{\tev}{\aunit{Te\kern -0.1em V}\xspace}
\newcommand{\gev}{\aunit{Ge\kern -0.1em V}\xspace}
\newcommand{\mev}{\aunit{Me\kern -0.1em V}\xspace}
\newcommand{\kev}{\aunit{ke\kern -0.1em V}\xspace}
\newcommand{\ev}{\aunit{e\kern -0.1em V}\xspace}
 
\newcommand{\mevc}{\ensuremath{\aunit{Me\kern -0.1em V\!/}c}\xspace}
\newcommand{\gevc}{\ensuremath{\aunit{Ge\kern -0.1em V\!/}c}\xspace}
\newcommand{\mevcc}{\ensuremath{\aunit{Me\kern -0.1em V\!/}c^2}\xspace}
\newcommand{\gevcc}{\ensuremath{\aunit{Ge\kern -0.1em V\!/}c^2}\xspace}
\newcommand{\gevgevcccc}{\ensuremath{\gev^2\!/c^4}\xspace} 

\def\fm   {\aunit{fm}\xspace}

\def\fb   {\ensuremath{\aunit{fb}}\xspace}
\def\invfb   {\ensuremath{\fb^{-1}}\xspace}

\def\gsim{{~\raise.15em\hbox{$>$}\kern-.85em
          \lower.35em\hbox{$\sim$}~}\xspace}
\def\lsim{{~\raise.15em\hbox{$<$}\kern-.85em
          \lower.35em\hbox{$\sim$}~}\xspace}

\def\pt         {\ensuremath{p_{\mathrm{T}}}\xspace}

\def\tell1  {TELL1\xspace}
\def\ukl1   {UKL1\xspace}

\newcommand{\phz}{\phantom{0}}
\newcommand{\lhcborcid}[1]{\href{https://orcid.org/#1}{\hspace*{0.1em}\raisebox{-0.45ex}{\includegraphics[width=1em]{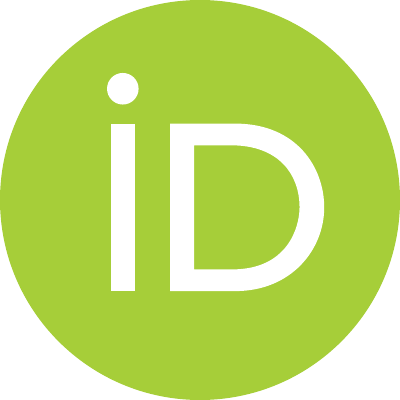}}}}

\hypersetup{
  colorlinks   = true, 
  urlcolor     = blue, 
  linkcolor    = blue, 
  citecolor    = red   
}

\ifEnableSectionTOCLinks
    \usepackage[explicit]{titlesec} 
    
    \let\oldcontentsline\contentsline
    \renewcommand

    \titleformat{\section}{\normalfont\Large\bf}{\hyperlink{tocsection.\thesection}{{\thesection} \parbox[t]{\dimexpr\textwidth-1pc}{#1}}}{1pc}{}

    \titleformat{\subsection}{\normalfont\bf}{\hyperlink{tocsubsection.\thesubsection}{{\thesubsection} \parbox[t]{\dimexpr\textwidth-1pc}{#1}}}{1pc}{}

    \titleformat{name=\section,numberless}[display]{}{}{0pt}{\normalfont\Huge\bfseries #1}
\fi

\usepackage{cite} 
\usepackage{LHCb/mciteplus}

\usepackage{longtable} 

\begin{document}

\renewcommand{\thefootnote}{\fnsymbol{footnote}}
\setcounter{footnote}{1}


\begin{titlepage}
\pagenumbering{roman}

\vspace*{-1.5cm}
\centerline{\large EUROPEAN ORGANIZATION FOR NUCLEAR RESEARCH (CERN)}
\vspace*{1.5cm}
\noindent
\begin{tabular*}{\linewidth}{lc@{\extracolsep{\fill}}r@{\extracolsep{0pt}}}
\ifthenelse{\boolean{pdflatex}}
{\vspace*{-1.5cm}\mbox{\!\!\!\includegraphics[width=.14\textwidth]{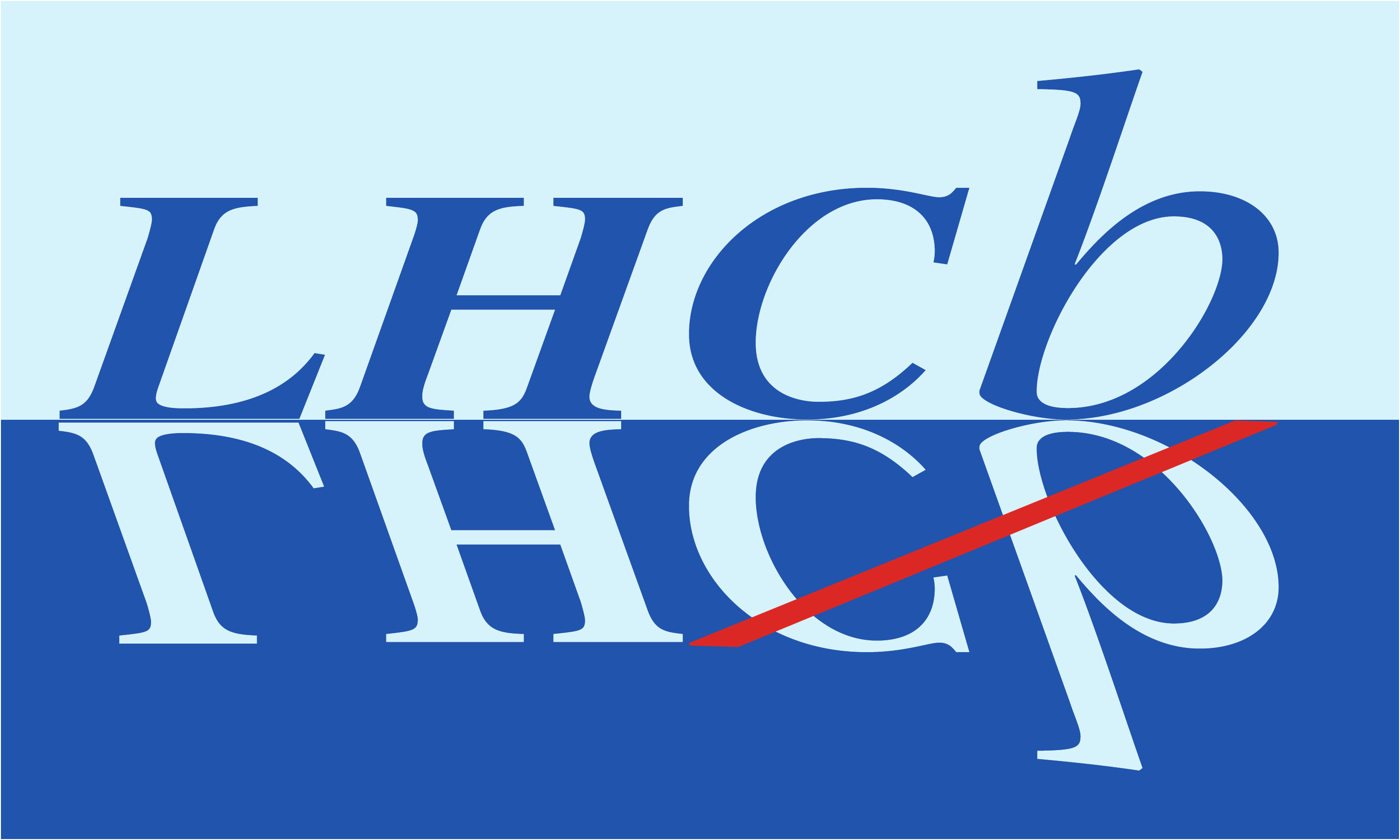}} & &}%
{\vspace*{-1.2cm}\mbox{\!\!\!\includegraphics[width=.12\textwidth]{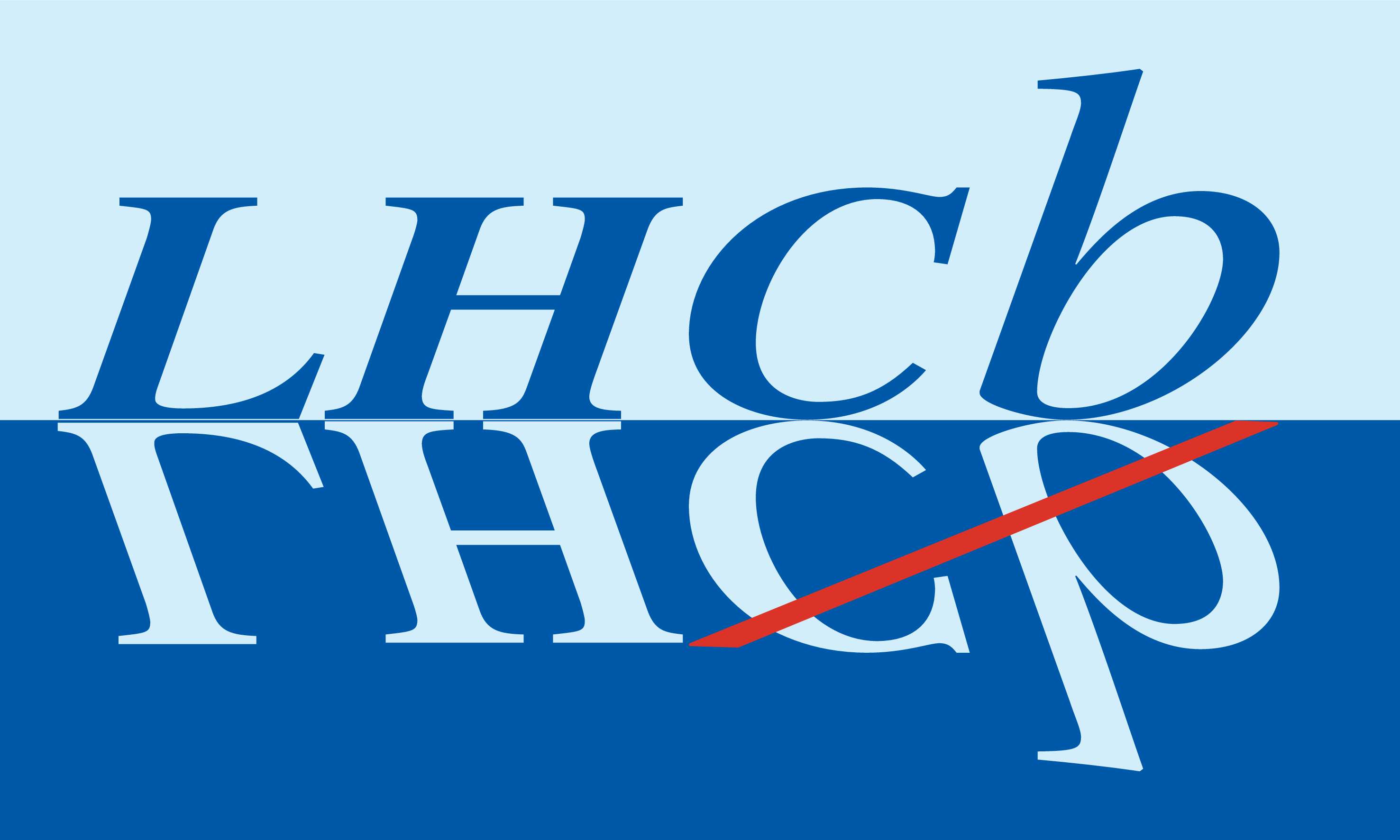}} & &}%
\\
 & & CERN-EP-2026-164 \\  
 & & LHCb-PAPER-2026-001 \\  
 & & 12 July 2026 \\ 
 & & \\
\end{tabular*}

\vspace*{1.0cm}

{\normalfont\bfseries\boldmath\huge
\begin{center}
  \papertitle 
\end{center}
}

\vspace*{2.0cm}

\begin{center}
\paperauthors\footnote{Authors are listed at the end of this Letter.}
\end{center}

\vspace{\fill}

\begin{abstract}
  \noindent
An amplitude analysis of the ${\ensuremath{\B^-}}\xspace\rightarrow {\ensuremath{\mathchar"7103^+_\ensuremath{c}}}\xspace\overline{p}\pi^{-}$ decay is performed using proton-proton collision data, corresponding to an integrated luminosity of $9\text{\,fb}^{-1}$, collected with the \mbox{LHCb} detector at center-of-mass energies of 7, 8, and 13$\text{\,Te\kern -0.1em V}$. 
In the $\ensuremath{\mathchar"7103^+_\ensuremath{c}}\pi^{-}$ invariant-mass spectrum, the known $\mathchar"7106_c(2455)^0$, $\mathchar"7106_c(2520)^0$, $\mathchar"7106_c(2800)^0$ baryons are observed, along with two new states: $\mathchar"7106_c(2900)^0$ and $\mathchar"7106_c(3200)^0$.
The masses and widths of the three high-mass states, modeled with \mbox{Breit--Wigner} line shapes, are determined. 
This result expands the landscape of charm baryon spectroscopy and provides crucial data for understanding charm baryon structure and dynamics.
\end{abstract}

\vspace*{2.0cm}

\begin{center}
  Submitted to Phys. Rev. Lett. 
\end{center}

\vspace{\fill}

{\footnotesize 
\centerline{\copyright~\papercopyright. \href{\paperlicenceurl}{\paperlicence}.}}
\vspace*{2mm}

\end{titlepage}


\newpage
\setcounter{page}{2}
\mbox{~}
%
%
%
%


\renewcommand{\thefootnote}{\arabic{footnote}}
\setcounter{footnote}{0}


\cleardoublepage


\pagestyle{plain} 
\setcounter{page}{1}
\pagenumbering{arabic}


Hadrons are composite systems of quarks bound by the strong interaction.
According to the quark model~\cite{quarkmodel1,quarkmodel2,Petermann:1965qlk}, hadrons are classified as mesons, with a minimum content of a quark and an antiquark, and baryons, built minimally from three quarks. Besides ground states, hadrons exhibit a spectrum of excited states resulting from distinct radial, spin or orbital angular momentum configurations.
The nonperturbative nature of low-energy QCD makes it difficult to calculate hadron properties, requiring the development of effective theoretical approaches for their study~\cite{Wang:2024jyk,Chen:2022asf}.

Baryons containing a single heavy quark are an ideal laboratory to understand the nature of low-energy strong interactions and verify theoretical models~\cite{Luo:2025sns}. 
The spectroscopy of singly heavy baryons is determined by the flavor and angular-momentum structure of the light quarks. 
The flavor structure of the two light quarks can be either symmetric or antisymmetric, which is the case for $\Lc$ and $\Sigmares_c$ baryons, respectively. Excited singly heavy baryons arise from radial and orbital excitations, either between the light quarks ($\rho$  mode) or between the light quark system and the heavy quark ($\lambda$ mode), combined with spin interactions, leading to a rich resonance spectrum. While several $\Lc$ baryons have been discovered, only few $\Sigmares_c$ hadrons are established~\cite{PDG2024}: the lower mass states $\Sigmares_c(2455)$ and $\Sigmares_c(2520)$ observed by the CLEO collaboration~\cite{CLEO:1996czm,CLEO:2000abd}, and the excited state $\Sigmares_c(2800)$ found by the \belle collaboration~\cite{Belle:2004zjl}. 
The $\Sigmares_c(2800)$ state is usually interpreted to be one of the seven 1P-wave $\lambda$-mode excitations, with masses predicted in the range $2.7$ to $2.9\gevcc$~\cite{Chen:2016spr,Cheng:2021qpd,WangWangKaiLei:2021kdd}. 
In 2008, the \babar collaboration reported the observation of these three $\Sigmares_c^0$ states in the $\Bm\to\Lc\antiproton\pim$ decay through a one-dimensional analysis of the $\Lc\pim$ mass spectrum~\cite{BaBar:2008get}.\footnote{The inclusion of charge-conjugated states and decays is implied throughout this Letter.} However, the measured $\Sigmares_c(2800)^0$ mass is inconsistent with the \belle analysis, hinting at a new state or a mixing of several states.
In the higher-mass region, 2S-, 1D- and 2P-wave excitations from both the $\rho$-mode and $\lambda$-mode would emerge according to calculations~\cite{Li:2024zze,Garcilazo:2007eh,Valcarce:2008dr,Jakhad:2023mni,Luo:2023sne,Yang:2021lce,Chen:2016iyi,Ebert:2011kk,Shah:2016mig}, demanding experimental verification. Measurements of masses and widths enable testing various theoretical calculations, and help to understand the dynamics of baryons containing both heavy and light quarks.

This Letter presents an amplitude analysis of the $\decay{\Bm}{\Lc\antiproton\pim}$ decay, where the $\Sigmares_c^0$ resonances are searched for in the $\Lc\pim$ final state.
The analysis is performed using proton-proton ($pp$) collision data collected by the~\lhcb experiment, corresponding to an integrated luminosity of $3\invfb$ at center-of-mass energies of 7 and 8\tev, and $6\invfb$ at 13\tev. 
The \lhcb detector is a
single-arm forward spectrometer covering the pseudorapidity range $2 < \eta < 5$ described in detail in Refs.~\cite{LHCb-DP-2008-001,LHCb-DP-2014-002}. Simulation, required to model the effects of the detector acceptance and the imposed selection requirements, is produced and corrected using data with software packages described in Refs.~\cite{Sjostrand:2007gs,LHCb-PROC-2010-056,Lange:2001uf,davidson2015photos,Allison:2006ve, *Agostinelli:2002hh,LHCb-PROC-2011-006,LHCb-PUB-2016-021,LHCb-DP-2013-002,LHCb-DP-2020-001,LHCb-DP-2018-001}.

The events used in this analysis are selected by topological triggers~\cite{LHCb-DP-2012-004,LHCb-DP-2019-001}. Offline, the $\Bm$ candidates are formed by combining $\Lc$, $\antiproton$ and $\pim$ candidates,
where the $\Lc$ baryons are reconstructed through the \decay{\Lc}{\proton\Km\pip} decay, with the invariant mass $m_{\proton\Km\pip}$ required to be consistent with the known $\Lc$ mass~\cite{PDG2024}.  
The final-state tracks are required to be identified as pions, kaons or protons by LHCb's ring imaging Cherenkov detectors, to have a large momentum transverse to the beam ($\pt$), and to be well separated from any primary collision vertex (PV).
The $\Bm$ and $\Lc$ candidates are required to have reconstructed decay vertices of good fit quality. 
In addition, the decay vertex of the $\Bm$ meson is required to be significantly displaced from its associated PV, defined as the PV
 that aligns best with the $\Bm$ flight direction.
To improve the experimental resolution on the reconstructed $\Bm$ invariant mass, $m_{\Lc\antiproton\pim}$, a kinematic fit to the whole decay chain is performed~\cite{Hulsbergen:2005pu},  with the $m_{\proton\Km\pip}$ invariant mass constrained to the known $\Lc$ mass~\cite{PDG2024} and the $\Bm$ momentum vector constrained to point back to its associated PV. 
The background, dominated by random combinations of $\Lc$, $\antiproton$, and $\pim$ hadrons,
is further suppressed by a boosted decision tree (BDT) classifier~\cite{Breiman,AdaBoost}, implemented in the TMVA toolkit~\cite{Hocker:2007ht,*TMVA4}. 
The classifier is trained using simulated \decay{\Bm}{\Lc\antiproton\pim} candidates as the signal, and candidates with invariant mass $m_{\Lc\antiproton\pim}$ exceeding $5.4 \gevcc$ from data as the background. 
Training variables for the BDT include those related to the decay topology, particle transverse momenta, vertex fit quality and particle identification (PID) information for pions, kaons and protons. 

\begin{figure}[!bp]
    \centering
    \begin{overpic}[width=0.49\columnwidth]{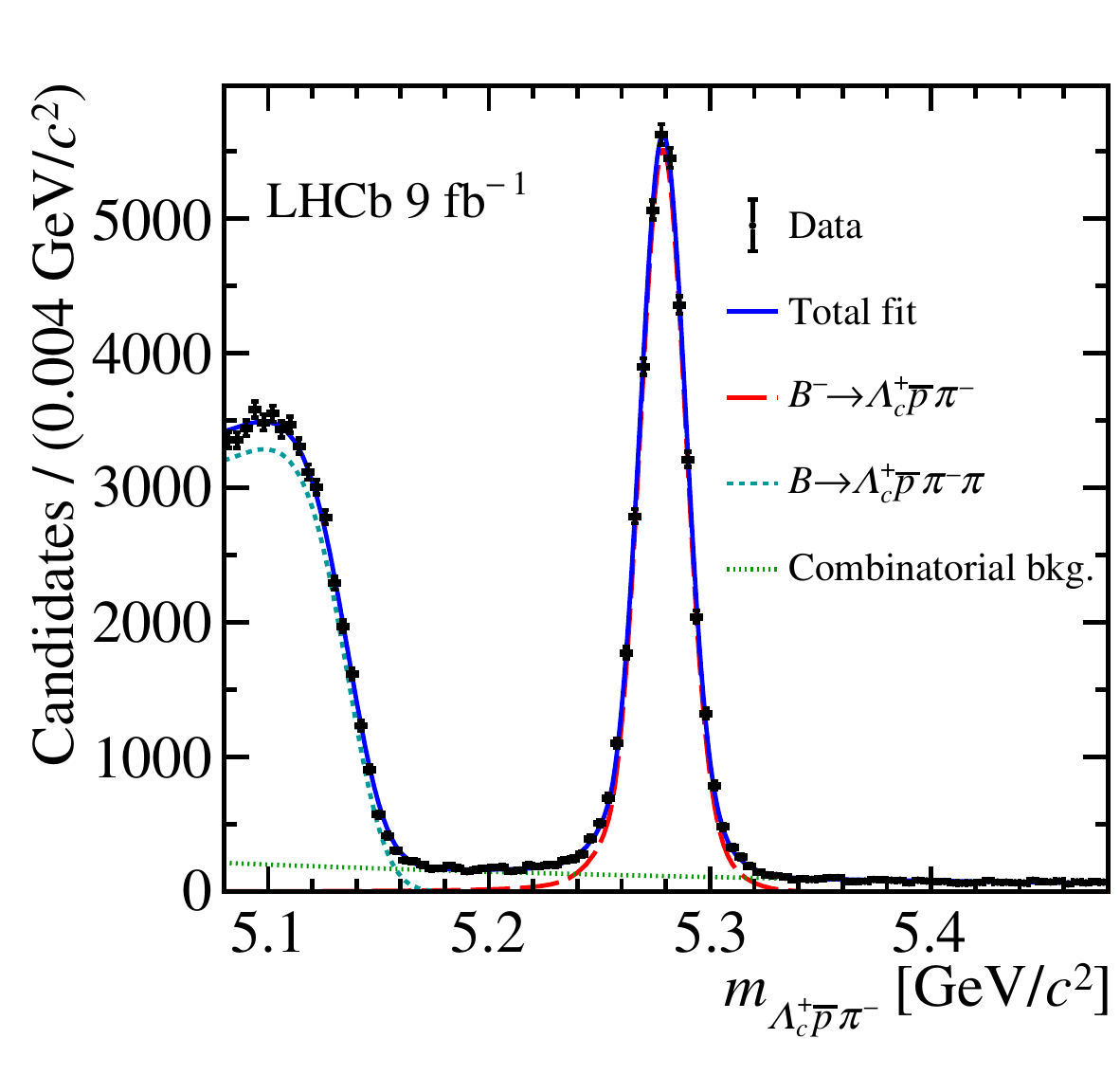}
    \put(85,80){(a)}
    \end{overpic}
    \begin{overpic}[width=0.49\columnwidth]{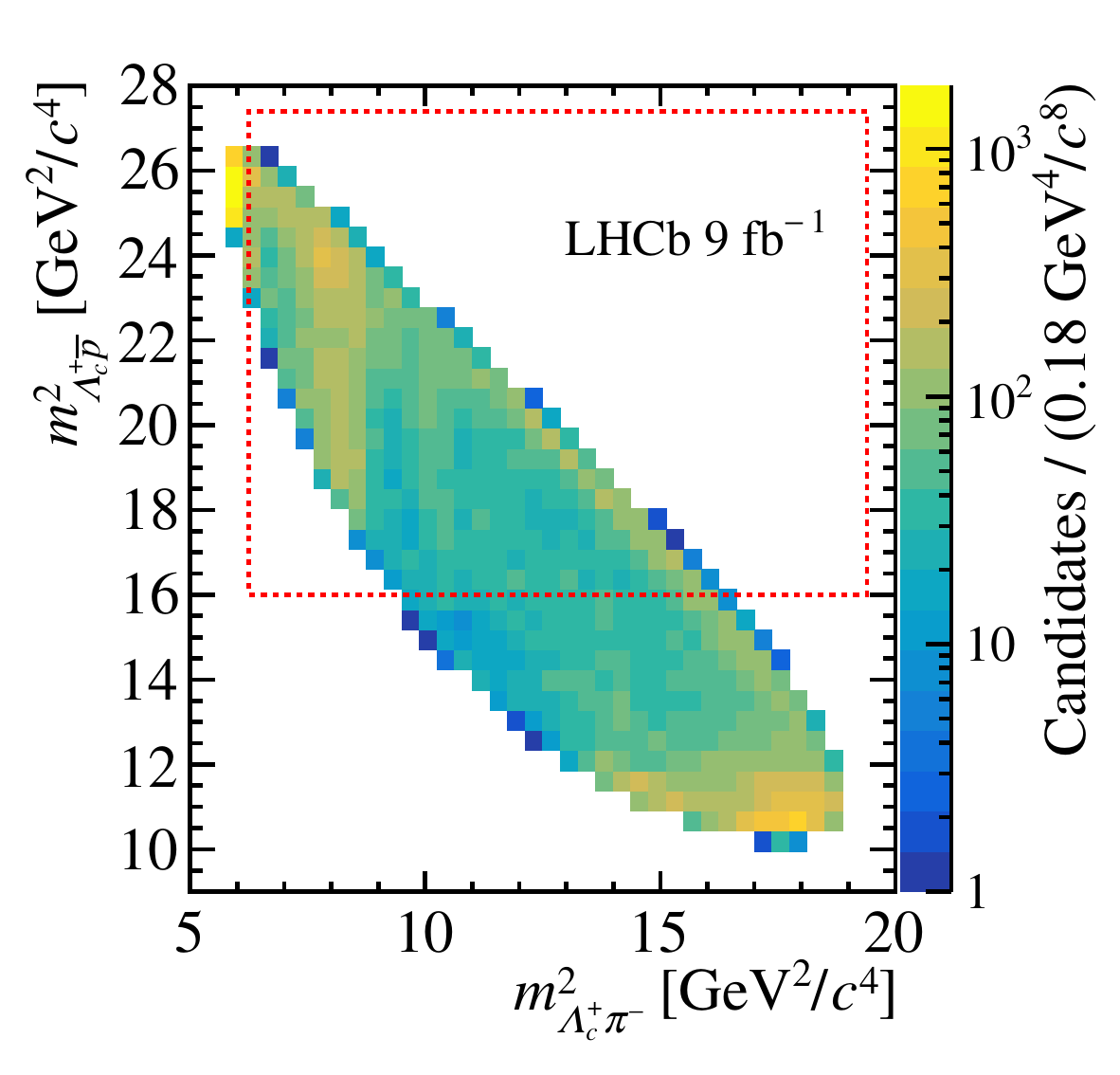}
    \put(70,80){(b)}
    \end{overpic}
    \caption{
    (left) Invariant-mass distribution of the ${\ensuremath{\mathchar"7103^+_\ensuremath{c}}}\xspace\overline{p}\pi^{-}$ system with the fit overlaid. 
    (right) Binned background-subtracted Dalitz plot of the $B^-$ candidates. Candidates with $m^2_{\ensuremath{\mathchar"7103^+_\ensuremath{c}}\pi^{-}}>6.25\ensuremath{\text{\,Ge\kern -0.1em V}^2\!/c^4}\xspace$ and $m^2_{\ensuremath{\mathchar"7103^+_\ensuremath{c}}\overline{p}}>16\ensuremath{\text{\,Ge\kern -0.1em V}^2\!/c^4}\xspace$ (inside the red dashed square) are considered for the amplitude fit. }
    \label{fig:massfit}
\end{figure}

In order to extract the signal yield, an unbinned maximum-likelihood fit is performed to the $m_{\Lc\antiproton\pim}$ distribution. Three components are identified in the distribution: the $\decay{\Bm}{\Lc\antiproton\pim}$ signal, the combinatorial background and the partially reconstructed \decay{\B}{\Lc \antiproton\pim\pi} decay, where the last pion is not reconstructed. 
The signal component is described by a combination of two Crystal Ball functions~\cite{Skwarnicki:1986xj} with a shared peak position.
The relative normalizations of the two Crystal Ball functions, the tail parameters and the ratio of their widths are fixed from simulation, while other parameters are floated in the fit. 
The combinatorial background is modeled by an exponential function. The partially reconstructed background is modeled using an ARGUS function convolved with a Gaussian function to account for experimental resolution~\cite{Argus:1990hfq}.
The $m_{\Lc\antiproton\pim}$ distribution is shown in Fig.~\ref{fig:massfit}(a), together with the fit results. 
In the signal region, $5.249<m_{\Lc\antiproton\pim}<5.309\gevcc$, approximately 
 $3$ times the  resolution around the known $\Bm$ mass~\cite{PDG2024}, 
the yield of the $\decay{\Bm}{\Lc\antiproton\pim}$ decay and combinatorial background fraction are determined to be 
$(37.0\pm 0.2)\times10^3$ and $f_\text{b}=(5.1\pm 0.5)\%$, respectively.

The Dalitz plot of the $\decay{\Bm}{\Lc\antiproton\pim}$ candidates in the signal region, represented in terms of the squared two-body invariant masses of the  \Lc\antiproton system, $m^2_{\Lc\antiproton}$, and \Lc\pim system, $m^2_{\Lc\pim}$, is shown in Fig.~\ref{fig:massfit}(b).
A few $\Sigmares_c^0$ states (collectively referred to as $\Sigmares_c^{**0}$) are clearly visible in the Dalitz plot as vertical bands,  
while $\Deltaresbar^{--}$ states (collectively referred to as $\Deltaresbar^{**--}$) are visible along the diagonal. 
In addition, a pronounced threshold enhancement is observed in the $\Lc \antiproton$ system for low $\Lc\antiproton$ invariant masses. This kind of enhancement is a common feature in baryon-antibaryon spectra~\cite{Chua:2001vh,Chua:2002wn}. 

A subsequent amplitude analysis is performed to candidates in the signal region to investigate resonant contributions to the $\decay{\Bm}{\Lc\antiproton\pim}$ decay.
Since the underlying mechanism responsible for the $\Lc \antiproton$ threshold enhancement is unclear, incorporating it into the amplitude fit presents significant challenges. 
To mitigate its potential impact on the analysis, the corresponding mass region $m^2_{\Lc \antiproton} < 16\gevgevcccc$ is excluded from the fit. 
Furthermore, since the narrow and overwhelming $\Sigmares_c(2455)^0$ resonance is well separated from other resonances in the Dalitz plot, leading to almost no interference between them, its contribution is excluded from the fit by requiring \mbox{$m^2_{\Lc\pim} > 6.25\gevgevcccc$}.
Instead, a separate amplitude fit is performed to candidates in the $\Sigmares_c(2455)^0$ resonance region, which supports $J=1/2$ for its spin, with detailed results shown in the End Matter. To reduce the complexity of the analysis, the $\Lc$ baryon is treated as a stable particle. In this case, the amplitude analysis is performed in the two-dimensional Dalitz plot of the $\decay{\Bm}{\Lc\antiproton\pim}$ three-body decay. 

The total probability density function (PDF) of the amplitude fit is defined as the sum of the signal and combinatorial background components, $\mathcal{P}_{\rm s}$ and $\mathcal{P}_{\rm b}$, respectively, with the background fraction $f_\text{b}$ fixed to the value obtained from the fit to the $m_{\Lc\antiproton\pim}$ mass spectrum (Fig.~\ref{fig:massfit}(a)). 
The combinatorial background PDF is determined using data from the sideband regions $5.175<m_{\Lc\antiproton\pim}<5.225\gevcc$ and $5.335<m_{\Lc\antiproton\pim}<5.385\gevcc$, assuming that the background behavior is the same in the signal region.
The signal PDF includes contributions from two interfering decay sequences, \decay{\Bm}{\Sigmares_c^{**0}(\to\Lc\pim) \antiproton} and \decay{\Bm}{\Deltaresbar^{**--}(\to\antiproton\pim)\Lc}, collectively denoted as \decay{\Bm}{R(\to a b) c}.
Various intermediate $\Sigmares_c^{**0}$ or $\Deltaresbar^{**--}$  states are investigated, each contributing with an independent amplitude, $\mathcal{M}_R(m,\theta|\vec{\omega})$. Here, $\theta$ denotes the angle between the $\Lc$ and $\Bm$ momentum vectors in the $R$ rest frame, and  $m$ is the invariant mass of the $\Lc\pim$ system. 
The vector of free parameters $\vec{\omega}$ includes the complex coupling of each amplitude, as well as the masses and widths of unestablished states.  
The distribution of candidates across the Dalitz plot is proportional to the PDF obtained by first constructing each spin-dependent decay amplitude as a coherent sum of the contributions from intermediate resonances, calculated within the helicity formalism~\cite{JACOB1959404}, and then incoherently summing the squared moduli of these amplitudes over the $\Lc$ and $\antiproton$ spin projections ($\lambda_c$ and $\lambda_p$)~\cite{multiplesolution}.
Accounting for the nonuniform experimental efficiency across the Dalitz plot, $\epsilon(m,\theta)$, the signal PDF is
\begin{equation}
        \mathcal{P}_\text{s}(m, \theta|\vec{\omega})=
      \frac{\epsilon(m,\theta)}{I(\vec{\omega})}\sum_{\lambda_c,\lambda_\proton}\left | \sum_R \mathcal{M}_{R,\lambda_c,\lambda_\proton}(m,\theta|\vec{\omega}) \right | ^2pq,
\end{equation}
where $I(\vec{\omega})$ is a normalization factor, $\lambda_c$ and $\lambda_p$ are helicities for $\Lc$ and $\antiproton$, $p$($q$) is the momentum of $a$($c$) in the $R$ rest frame, and the index $R$ runs over the considered contributions.
The efficiency is determined in bins of the Dalitz plot using simulated \decay{\Bm}{\Lc\antiproton\pim} decays.
The log-likelihood is built as
\begin{equation}
 \ln\mathcal{L}=\sum_{j}\ln\left[ (1-f_\text{b})\mathcal{P}_\text{s}(m^j,\theta^j|\vec{\omega}) +f_\text{b}\mathcal{P}_\text{b}(m^j,\theta^j) \right],
\end{equation}
where the index $j$ runs over the number of $\Bm$ candidates in the signal region.

The signal model includes $\Sigmares_c^{**0}$ and $\Deltaresbar^{**--}$ contributions in the $\Lc\pim$ and $\antiproton\pim$ channels, respectively, and a nonresonant (NR) component in the $\Lc\pim$ channel. The currently known states, $\Sigmares_c(2520)^0$, $\Sigmares_c(2800)^0$, $\Deltaresbar(1232)^{--}$, $\Deltaresbar(1600)^{--}$, $\Deltaresbar(1620)^{--}$ and $\Deltaresbar(1700)^{--}$, are initially included in the fit. 
Since there is a clear structure around $3.2\gevcc$ in $m_{\Lc\pim}$ and the bump around $2.8\gevcc$ is asymmetric in the $\Lc\pim$ spectrum, two additional $\Sigmares_c^{**0}$ states, $\Sigmares_c(2900)^0$ and $\Sigmares_c(3200)^0$, are found to be needed to describe data.
For the $\Sigmares_c^{**0}$ states, the spin-parity ($J^P$) quantum numbers are not known, and therefore six possible hypotheses $1/2^{\pm}$, $3/2^{\pm}$ and $5/2^{\pm}$ are tested in the fit.
For the $\Sigmares_c(2520)^0$ baryon, the spin-parity is measured to be $3/2^+$ with a significance exceeding 6 standard deviations ($\sigma$) relative to other hypotheses. This result is consistent with the quark-model expectations~\cite{PDG2024}.
The $\Deltaresbar(1620)^{--}$ component has a significance below $1\sigma$ 
and is therefore not included in the baseline model.
The mass-dependent lineshape of each resonance is modeled using a relativistic Breit--Wigner function, where the masses and widths are floated for the three $\Sigmares_c^{**0}$ states with mass larger than $2.6\gevcc$, while they are fixed to their known values~\cite{PDG2024} for the remaining states. Blatt--Weisskopf form factors are used to model the angular momentum barrier, where the resonance size is set to $4~(\!\gevc)^{-1} \approx 0.8\fm $~\cite{Back:2017zqt}.
An S-wave NR contribution is also needed to model the $m_{\Lc\pim}$ distribution. This mass-dependent lineshape is described by an exponential function, where the slope is allowed to vary. 
The contribution of a P-wave NR component is also tested but found to be negligible and therefore not considered in the subsequent studies.

\begin{figure*}[!t]
    \centering
    \includegraphics[width=0.32\columnwidth]{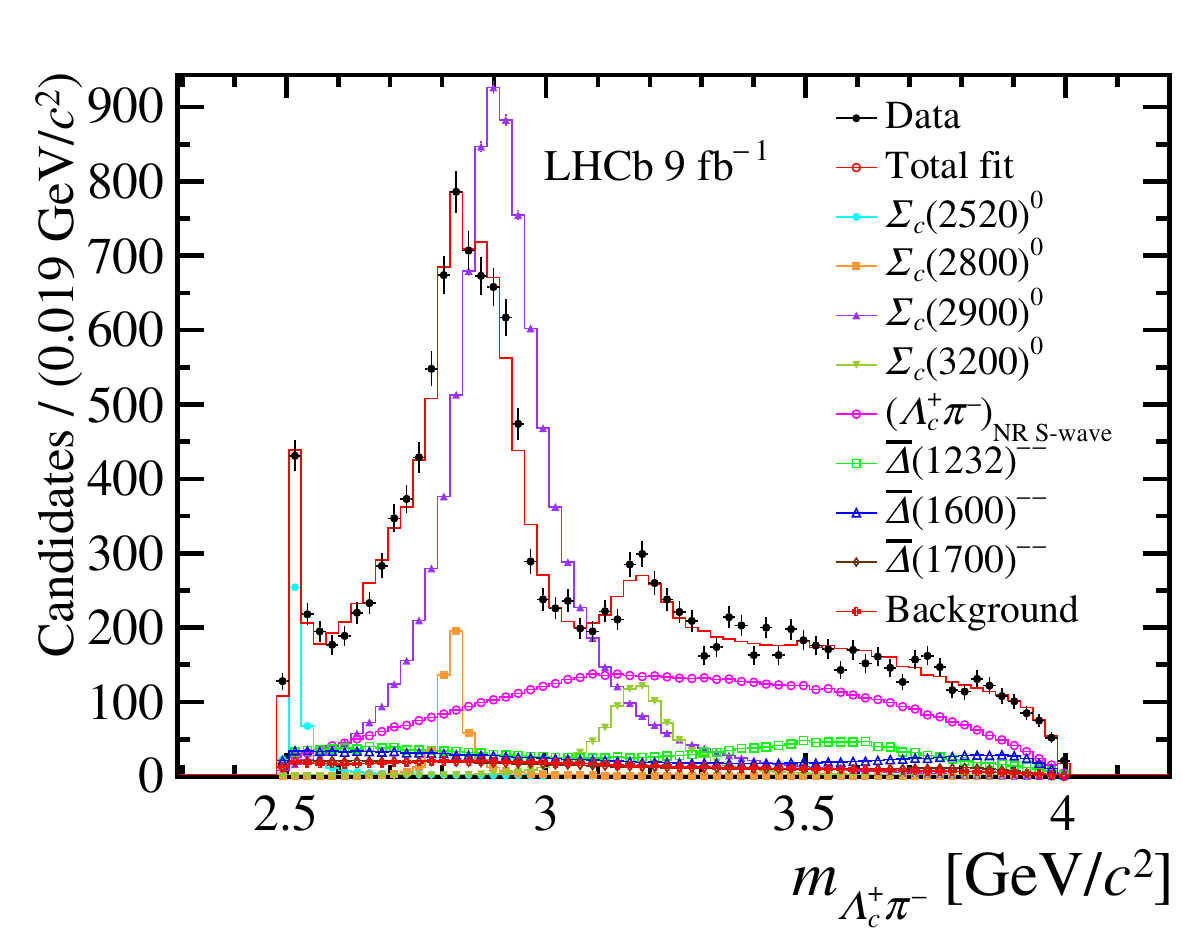}
    \includegraphics[width=0.32\columnwidth]{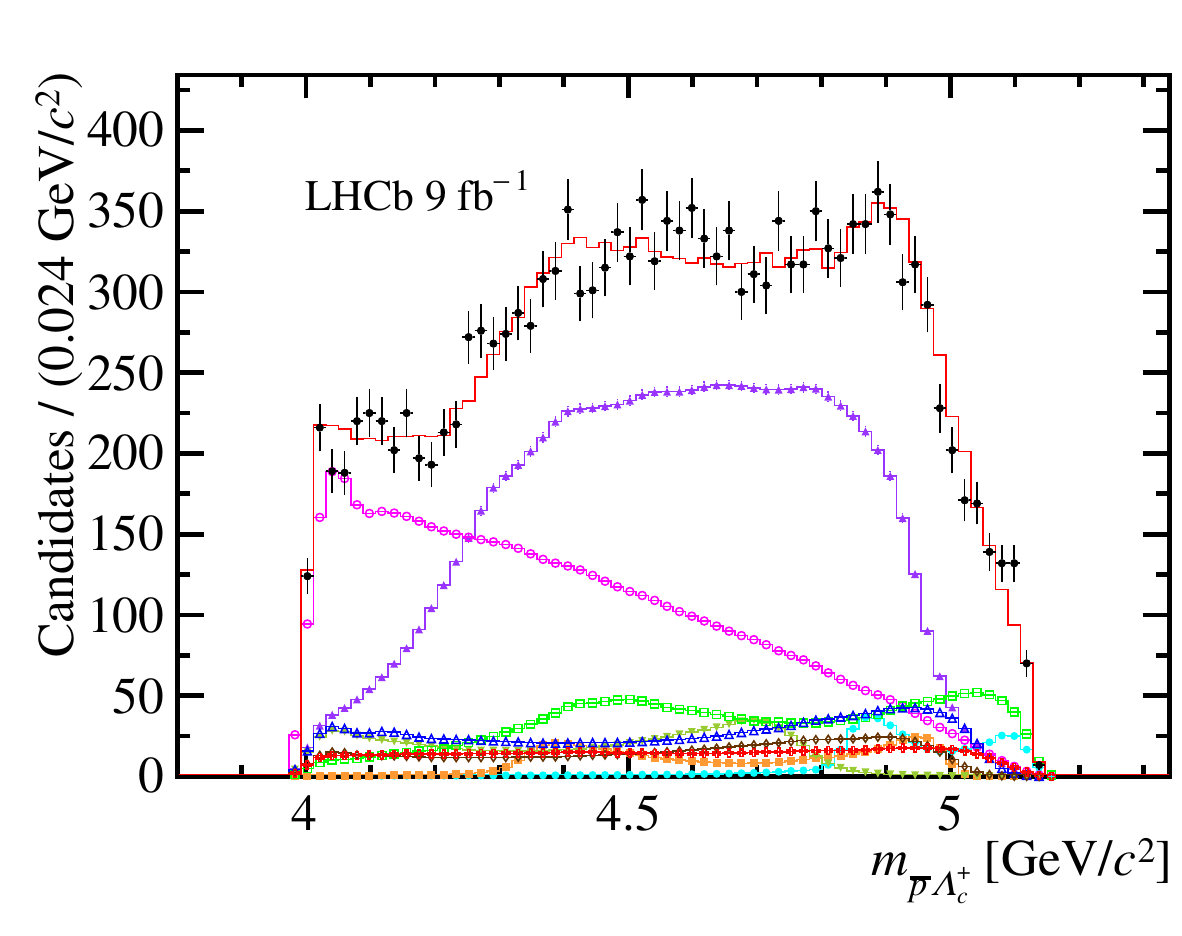}
    \includegraphics[width=0.32\columnwidth]{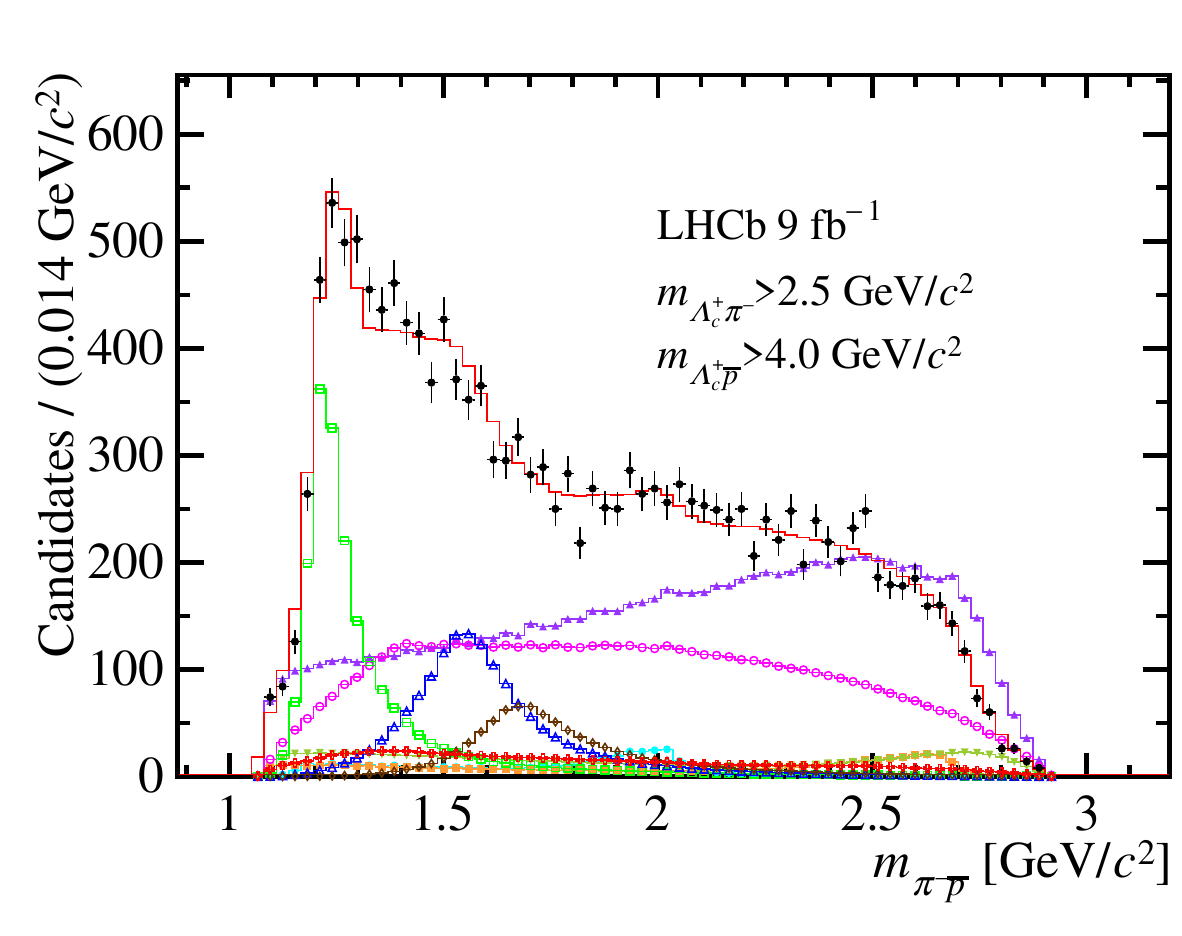}
    \includegraphics[width=0.32\columnwidth]{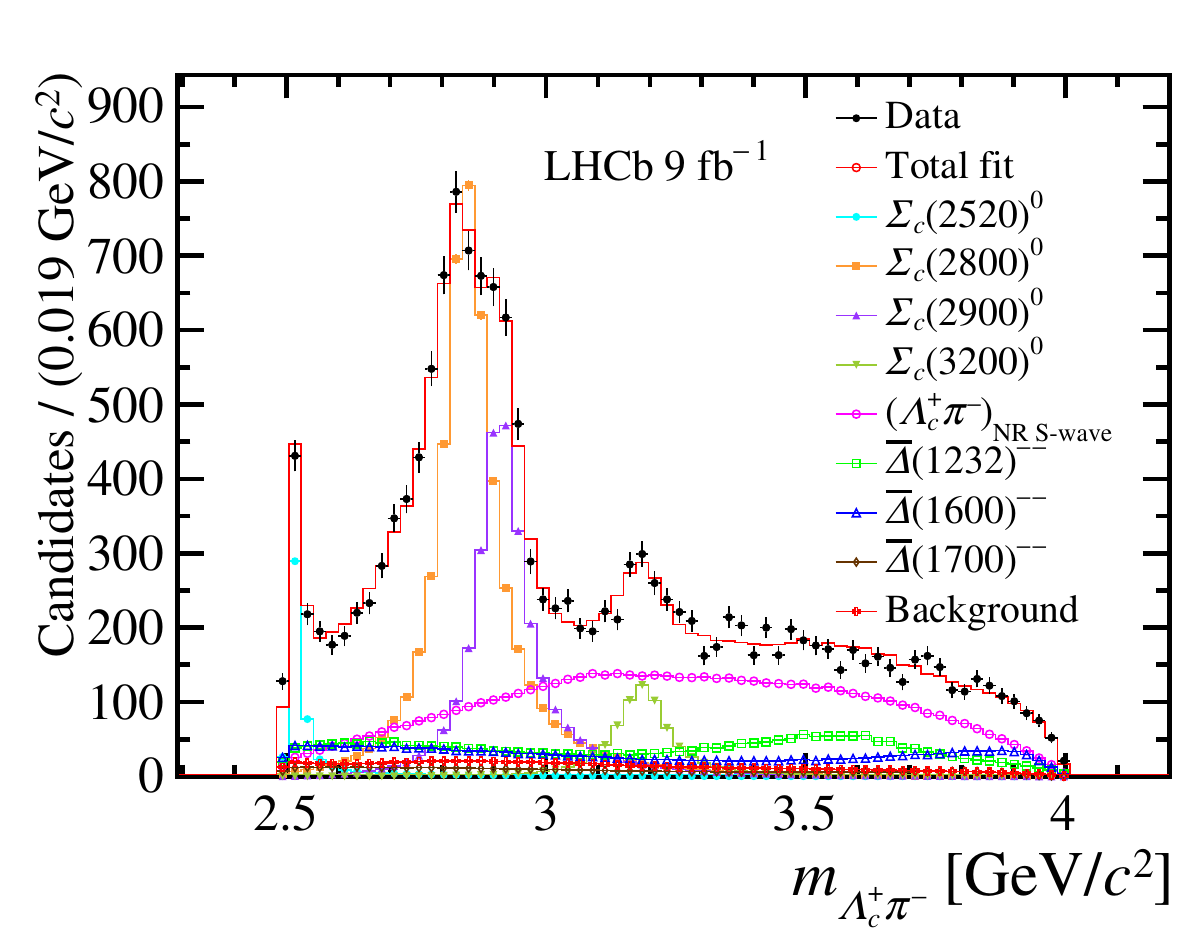}
    \includegraphics[width=0.32\columnwidth]{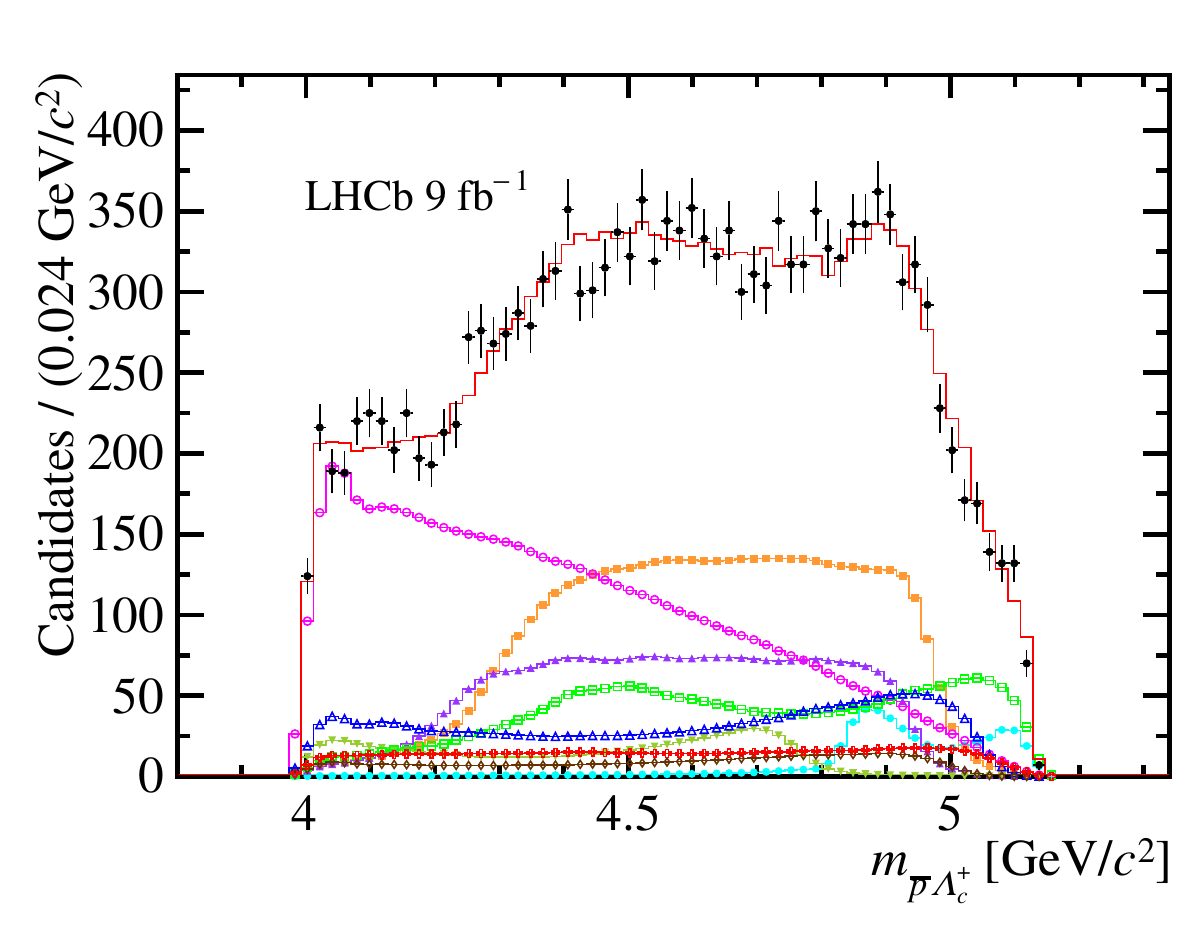}
    \includegraphics[width=0.32\columnwidth]{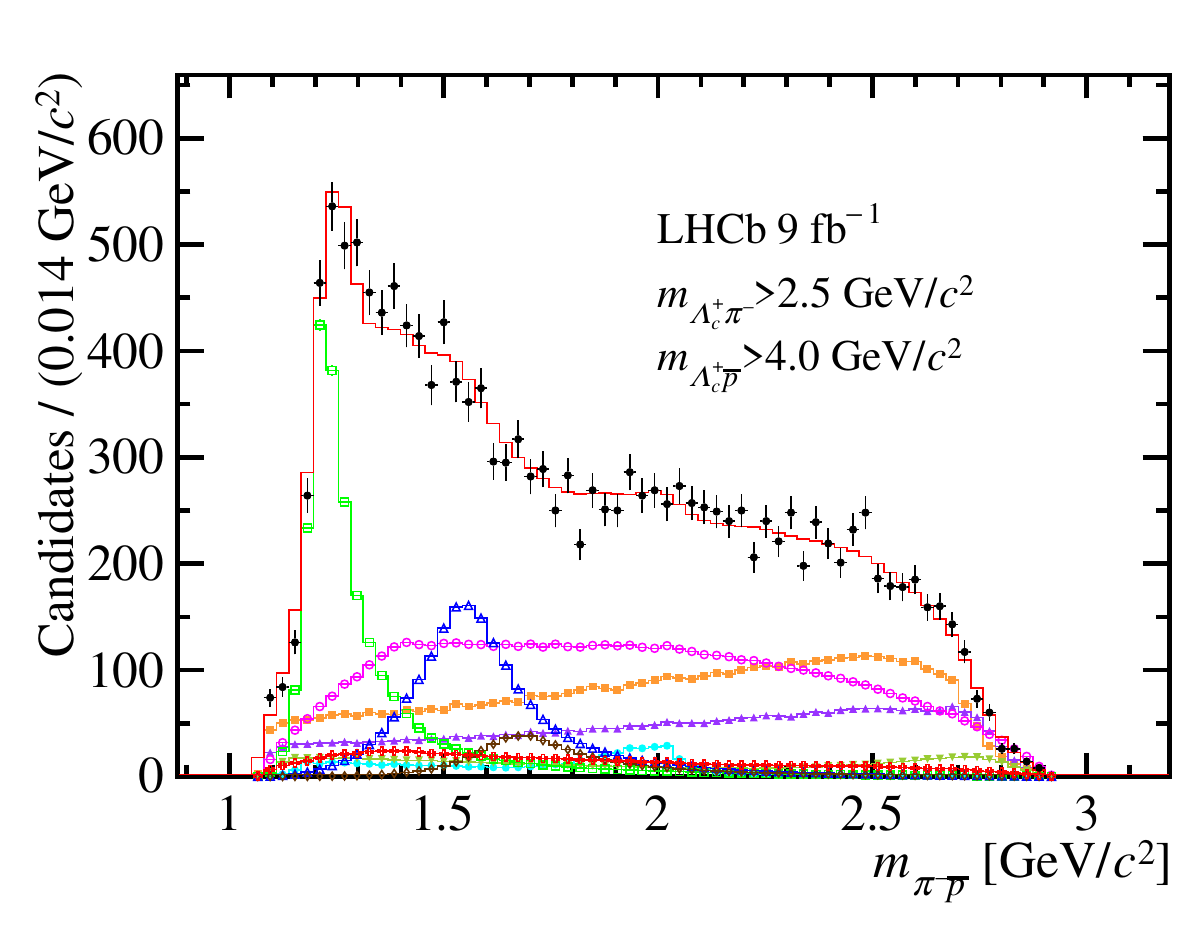}
    \caption{Distributions of (left) $m_{\mathchar"7103_c^+\pi^{-}}$, (middle) $m_{\mathchar"7103_c^+\overline{p}}$ and (right) $m_{\overline{p}\pi^{-}}$ invariant masses. The results of the baseline fit for solutions of groups (top) A and (bottom) B are also shown.
    }
    \label{fig:amp_fit}
\end{figure*}

The fit is performed by minimizing the negative log-likelihood (NLL). However, solutions of multiple $J^P$ configurations yield similar minimum NLL values, 
preventing an unambiguous determination of the spin-parity quantum numbers.
Therefore, the solutions with twice the likelihood difference $2\Delta\ln\mathcal{L} < 25$ relative to the fit with smallest NLL are grouped into two categories (A and B) according to the fit results, which are similar within each group but different between the two groups. The two sets of solutions are reported separately.
The configuration yielding the smallest NLL includes three $\Sigmares_c^{**0}$ states: $\Sigmares_{c}(2800)^0$, $\Sigmares_{c}(2900)^0$, and $\Sigmares_{c}(3200)^0$, with spin-parity assignments $J^P = 3/2^+$, $1/2^-$, and $3/2^-$, respectively. This solution, together with other $J^P$ configurations that yield similar masses and widths for the $\Sigmares_c^{**0}$ states, is classified as group A. For solutions in group B, the measurements of the mass and width of the $\Sigmares_{c}(2800)^0$ state and the width of the $\Sigmares_{c}(2900)^0$ state differ significantly from those  in group A, while those for the $\Sigmares_c(3200)^0$ state are compatible.
Among the spin-parity assignments for solutions in group B, that with $J^P = 1/2^-$, $1/2^+$, and $3/2^-$ for the three $\Sigmares_c^{**0}$ states yields the smallest NLL. 
The likelihood difference  $2\Delta\ln\mathcal{L}$ between the best-fit solutions in groups A and B is less than 1, indicating that the two solution families are statistically indistinguishable.
The two-body invariant-mass distributions and projections of the amplitude fit are shown in Fig.~\ref{fig:amp_fit}, where each component is normalized according to its fraction ($\mathcal{F}$), calculated as the ratio of the phase-space integral of the squared amplitude for that resonance to the integral of the total squared amplitude. Despite the differences in the individual contributions between the two groups, the overall fits describe the data equally well. The measured masses and widths of the three $\Sigmares_c^{**0}$ states are listed in Table~\ref{table:finalresults_s1}, and the fractions are listed in Table~\ref{tab:fitfraction},  where the quoted statistical uncertainties are obtained using pseudoexperiments.

The best solution in each group is taken as baseline model, while the spread among solutions within the same group is assigned as a source of systematic uncertainty.
The significance of the three $\Sigmares_c^{**0}$ states is evaluated using likelihood-ratio tests based on pseudoexperiments~\cite{LHCb-PAPER-2020-025}. 
The two-resonance hypothesis to model the structure at $2.8\gevcc$, corresponding to the $\Sigmares_c(2800)^0$ and $\Sigmares_c(2900)^0$ states, is favored over the single-resonance hypothesis with a significance of $7.5\sigma$, including systematic uncertainties.
The $\Sigmares_c(3200)^0$ state is observed with a significance of $12.0 \sigma$. Additional resonant contributions in the $\Lc\pim$ system are investigated but found to be negligible.

\begin{table}[bt]
\centering
\caption{Masses, widths and relative fit fractions ($\mathcal{R}(\Sigmares_c^{**0})$) of $\decay{\Bm}{\Sigmares_c^{**0}(\to \Lc\pim)\antiproton}$ relative to $\decay{\Bm}{\Sigmares_c(2455)^0(\to \Lc\pim)\antiproton}$  for the two groups of solutions, where the first uncertainties are statistical and the second are systematic. The $J^P$ of the $\Sigmares_c(2520)^{0}$ state is measured to be $3/2^+$.}
\label{table:finalresults_s1}
\renewcommand{\arraystretch}{1.2}
\begin{tabular}{lll}
\hline
Parameter &   \phantom{XXXX}Group A & \phantom{XXXX}Group B\\
\hline
$m_{\Sigmares_c(2800)^0}[\!\gevcc]     $&$ 2.819 \pm 0.006 \pm 0.002  $ & $2.848\pm0.004 \pm 0.006  $\\ 
$\Gamma_{\Sigmares_c(2800)^0}[\!\gev] $&$ 0.033\pm0.007 \pm 0.008      $ & $0.099\pm0.007 \pm 0.021     $\\
$m_{\Sigmares_c(2900)^0}[\!\gevcc]      $&$ 2.908\pm0.005 \pm 0.009  $ & $2.914\pm0.003 \pm 0.008  $\\ 
$\Gamma_{\Sigmares_c(2900)^0}[\!\gev] $&$0.175\pm0.008 \pm 0.023     $ & $0.092\pm0.006 \pm 0.023     $\\
$m_{\Sigmares_c(3200)^0}[\!\gevcc]      $&$ 3.186\pm0.006 \pm 0.014 $ & $3.190\pm0.005 \pm 0.011 $\\
$\Gamma_{\Sigmares_c(3200)^0}[\!\gev] $&$ 0.133\pm0.018 \pm 0.027   $ & $0.100\pm0.015 \pm 0.032   $\\
$\mathcal{R}$($\Sigmares_{c}(2520)^0$) & $\phantom{00}0.09\pm0.01 \pm 0.02$ & $\phantom{00}0.11\pm0.01 \pm 0.03$\\
$\mathcal{R}$($\Sigmares_{c}(2800)^0$) & $\phantom{00}0.09\pm0.02 \pm 0.05$ & $\phantom{00}0.86\pm0.07 \pm 0.42$\\
$\mathcal{R}$($\Sigmares_{c}(2900)^0$) & $\phantom{00}1.68\pm0.07 \pm 0.29$ & $\phantom{00}0.50\pm0.04 \pm 0.19$\\
$\mathcal{R}$($\Sigmares_{c}(3200)^0$) & $\phantom{00}0.21\pm0.03 \pm 0.05$ & $\phantom{00}0.17\pm0.03 \pm 0.05$\\
$\mathcal{R}$($(\Lc\pim)_{\rm NR\, S-wave}$) & $\phantom{00}2.19\pm0.14 \pm 0.22$ & $\phantom{00}2.23\pm0.11 \pm 0.30$\\
\hline
\end{tabular}
\end{table}

The ratio between the fractional contribution to \decay{\Bm}{\Lc\antiproton\pim} decay of the high-mass $\Sigmares_c^{**0}$ resonance and that of the $\Sigmares_c(2455)^0$ resonance is determined as
\begin{equation}
\mathcal{R}(\Sigmares_c^{**0}) \equiv \dfrac{\mathcal{F}_{\text{DP}}[\decay{\Bm}{\Sigmares_c^{**0}(\to\Lc\pim)\antiproton}]}{\mathcal{F}_{\text{DP}}
[\decay{\Bm}{\Sigmares_c(2455)^0(\to\Lc\pim)\antiproton}]},
\end{equation}
where the variable $\mathcal{F}_{\text{DP}}$ is the fraction of each amplitude contribution determined for the full Dalitz plot region.

Various systematic uncertainties are considered for the masses, widths and relative fit fractions of the $\Sigmares_c^{**0}$ states, as summarized in Tables~\ref{tab:syserrA} and~\ref{tab:syserrB} in the End Matter. 
The uncertainty associated with the invariant-mass fit is evaluated by repeating the amplitude fit while varying the background fraction within its statistical uncertainty and by modifying the mass-fit parametrization.
Efficiency uncertainties across the Dalitz plot arise from several sources: the binning scheme of the efficiency map, assessed using alternative binning schemes; limited simulation sample sizes, estimated by resampling of the efficiency map;  imperfections in track-efficiency and modeling of the hardware-level trigger responses, corrected using the data~\cite{LHCb-DP-2020-001,LHCb-DP-2013-002,LHCb-DP-2018-001}; and the discrepancy between the data and simulated samples on variables used in the BDT training, corrected by weighting simulated samples to data. 
The impact of the $\Lc$ decay is accounted for by weighting simulation samples to match data. Since the weighting depends on the chosen phase space, different phase-space regions are used to assess this effect, and the largest resulting differences are taken as a systematic uncertainty.
The influence of the $\Lc$ polarization is also considered by comparing data and simulation for distributions related to the $\Lc$ polarization and is found to be negligible.
Uncertainties arising from fixed parameters are also taken into account, including the resonance radii and the $\Deltaresbar^{**--}$ resonance parameters. The uncertainty from the resonance radii is studied by varying the baseline values of 4\,$(\!\gevc)^{-1}$ to 3 or 5\,$(\!\gevc)^{-1}$. The uncertainty associated with fixing the $\Deltaresbar^{**--}$ masses and widths is evaluated by allowing these parameters to float in the amplitude fit, with Gaussian constraints centered on their known values.
For uncertainties associated with alternative models, instead of a coherent sum of Breit--Wigner amplitudes, K-matrix models~\cite{Chung:1995dx} are used to describe the $\Deltaresbar^{**--}$ contributions with $J^P=3/2^-$ and $\Lc\pim$ contributions with $J^P=1/2^-$, respectively. 
Different nonresonant models, such as second-order polynomial functions, are tested. 
Since the $J^P$ quantum numbers of the three $\Sigmares_c^{**0}$ states cannot be determined unambiguously, the spread among solutions within each group is treated as a source of systematic uncertainty, labeled as the $J^P$ configurations. The systematic uncertainty associated with the possible contribution of a $\Lc\antiproton$ resonance is evaluated by tightening the selection on $m_{\Lc\antiproton}$, which has a negligible influence.
The total systematic uncertainties are determined by combining all contributions in quadrature.

According to the results summarized in Table~\ref{table:finalresults_s1}, in group A two states, one relatively narrow and one broader, are present in the $2.8$\gevcc mass region,  while in group B, the two $\Sigmares_c^{**0}$ states have similar width. 
Fit fractions of the three high-mass excited $\Sigmares_c^0$ states to the \decay{\Bm}{\Lc\antiproton\pim} decay are found to be comparable to that of the $\Sigmares_c(2455)^0$ baryon.
Compared with previous measurements, the $\Sigmares_c(2800)^0$ state reported by the \babar collaboration in the \decay{\Bm}{\Lc\antiproton\pim} decay is resolved into two states in this analysis.
The mass and width of the $\Sigmares_c(2800)^0$ baryon in group A are consistent with those reported by the \belle collaboration in inclusive production~\cite{Belle:2004zjl}.
The measured properties of the $\Sigmares_c(3200)^0$ state are consistent between groups A and B, allowing the results to be combined. The group A results are taken as baseline. The systematic uncertainty is taken as the larger one in groups A and B. The difference between the results in groups A and B is assigned as an additional systematic uncertainty. As a result, the $\Sigmares_c(3200)^0$ mass, width and relative fraction are measured to be \mbox{$m_{\Sigmares_c(3200)^0}=3.186\pm0.006 \pm 0.015\gevcc$}, \mbox{$\Gamma_{\Sigmares_c(3200)^0}=0.133\pm0.018 \pm 0.046\gevcc$} and \mbox{$\mathcal{R}(\Sigmares_{c}(3200)^0)=0.21\pm0.03 \pm 0.06$}.
According to theoretical predictions, several $\Sigmares_c^0$ states are expected to populate the mass region explored in this analysis. The $\Sigmares_c(2800)^0$ state could be interpreted as an excited 1P- or a 2S-wave  state~\cite{Chen:2016iyi,Ebert:2011kk,Shah:2016mig,Capstick:1986ter,Ebert:2007nw}. The newly observed $\Sigmares_c(2900)^0$ state may be an excited 2S- or 1P-wave state~\cite{Ebert:2011kk,Capstick:1986ter}. The newly observed $\Sigmares_c(3200)^0$ state may correspond to either an excited 2P- or 3S-wave state~\cite{Ebert:2011kk, Shah:2016mig, Capstick:1986ter}. 
Further experimental measurements, including other decay channels, will help to shed further light on the nature of these states.

In conclusion, an amplitude analysis of the \decay{\Bm}{\Lc\antiproton\pim} decay is performed using $pp$ collisions at center-of-mass energies of $7, 8$ and $13\tev$ collected by the \lhcb detector, corresponding to an integrated luminosity of $9\invfb$. 
In addition to the $\Sigmares_c(2455)^0$ and 
$\Sigmares_c(2520)^0$ states, three other excited $\Sigmares_c^{**0}$ states, $\Sigmares_c(2800)^0$, $\Sigmares_c(2900)^0$ and $\Sigmares_c(3200)^0$, are needed to describe the $\Lc \pim$ mass spectrum. The spin of the $\Sigmares_c(2455)^0$ state is measured to be $1/2$ and the spin-parity of the $\Sigmares_c(2520)^0$ state is determined to be $3/2^+$. 
Two distinct solutions are obtained from the amplitude fit. 
The masses, widths, and fit fractions of the $\Sigmares_c(2800)^0$ and $\Sigmares_c(2900)^0$ states exhibit significant differences between the two solutions, whereas the corresponding parameters for the $\Sigmares_c(3200)^0$ state are compatible.
The $\Sigmares_c(2800)^0$ state reported by the \babar collaboration in the decay \decay{\Bm}{\Lc\antiproton\pim} is found to be consistent with a superposition of two overlapping resonances. The mass and width of the $\Sigmares_c(2800)^0$ from solution group A are consistent with the \belle measurement, which does not find evidence for the $\Sigmares_c(2900)^0$ contribution, while the results from solution group B show discrepancies with the \belle results.
The new $\Sigmares_c(2900)^0$ and $\Sigmares_c(3200)^0$ states represent the first observation of excited $\Sigmares_c^0$ states in over two decades.
This analysis significantly extends the spectroscopy of $\Sigmares_c^0$ states and helps to understand the internal structure and low-energy dynamics of baryons containing a heavy quark.


\vspace{1cm}

\section*{Acknowledgements}
%
%
\noindent We express our gratitude to our colleagues in the CERN
accelerator departments for the excellent performance of the LHC. We
thank the technical and administrative staff at the LHCb
institutes.
We acknowledge support from CERN and from the national agencies:
ARC (Australia);
CAPES, CNPq, FAPERJ and FINEP (Brazil); 
MOST and NSFC (China); 
CNRS/IN2P3 and CEA (France);  
BMFTR, DFG and MPG (Germany);
NKFIH (Hungary);              
INFN (Italy); 
NWO (Netherlands); 
MNiSW and NCN (Poland); 
MEC/IFA (Romania); 
MICIU and AEI (Spain);
SNSF and SER (Switzerland); 
NASU (Ukraine); 
STFC (United Kingdom); 
DOE NP and NSF (USA).
We acknowledge the computing resources that are provided by ARDC (Australia), 
CBPF (Brazil),
CERN, 
IHEP and LZU (China),
IN2P3 (France), 
KIT and DESY (Germany), 
INFN (Italy), 
SURF (Netherlands),
Polish WLCG (Poland),
IFIN-HH (Romania), 
PIC (Spain), CSCS (Switzerland), 
GridPP (United Kingdom),
and NSF (USA).  
We are indebted to the communities behind the multiple open-source
software packages on which we depend.
Individual groups or members have received support from
RTP (Australia), 
FWO Odysseus grant G0ASD25N (Belgium), 
Key Research Program of Frontier Sciences of CAS, CAS PIFI, CAS CCEPP (China); 
Minciencias (Colombia);
EPLANET, Marie Sk\l{}odowska-Curie Actions, ERC and NextGenerationEU (European Union);
A*MIDEX, ANR, IPhU and Labex P2IO, and R\'{e}gion Auvergne-Rh\^{o}ne-Alpes (France);
Alexander-von-Humboldt Foundation (Germany);
ICSC (Italy); 
Severo Ochoa and Mar\'ia de Maeztu Units of Excellence, GVA, XuntaGal, GENCAT, InTalent-Inditex and Prog.~Atracci\'on Talento CM (Spain);
the Leverhulme Trust, the Royal Society and UKRI (United Kingdom).



\clearpage
\section*{End Matter}

\subsection*{Amplitude fit in the \texorpdfstring{\boldmath{$\Sigmares_c(2455)^0$}}{Sigmac(2455)0} mass region}
\begin{figure*}[!htbp]
    \centering
    \begin{overpic}[scale=0.35]{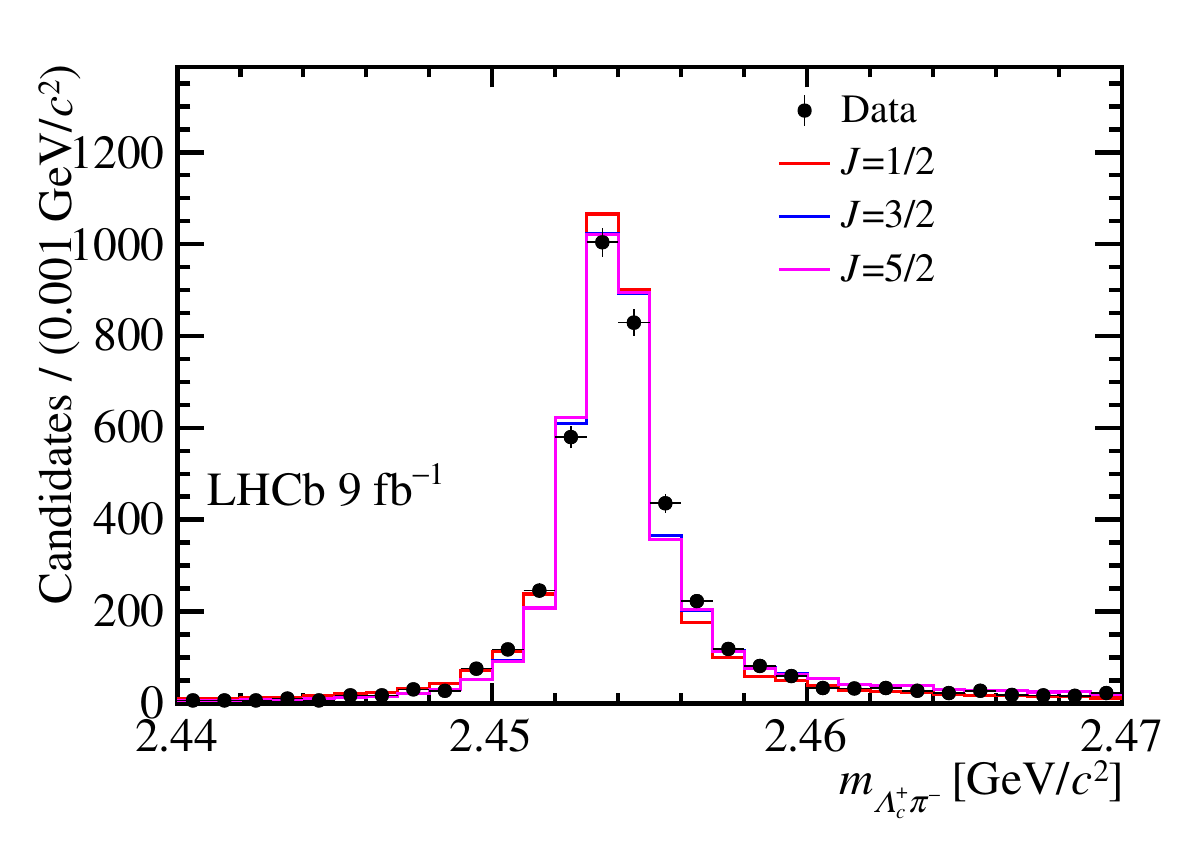}
     \put(18,60){(a)}   
    \end{overpic}
    \begin{overpic}[scale=0.35]{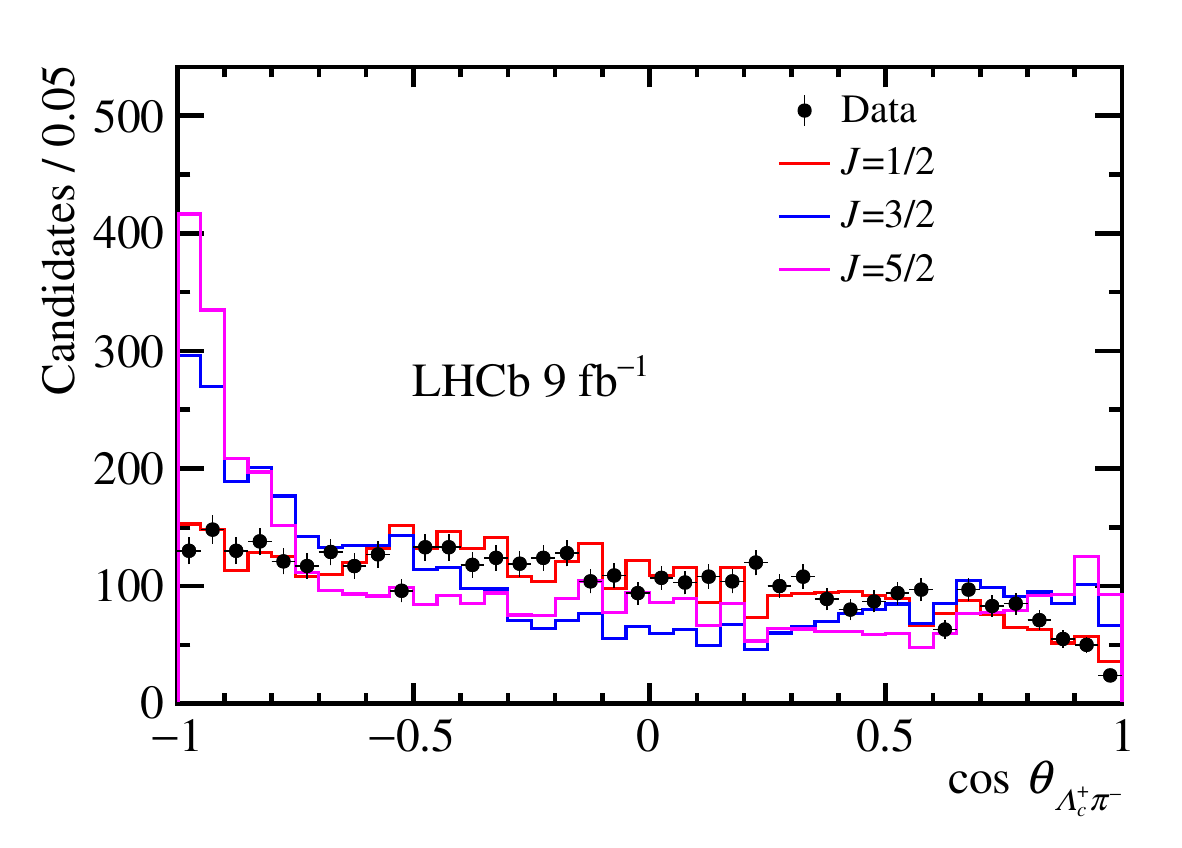}
     \put(18,60){(b)}   
    \end{overpic}
    \caption{Distributions of the (a) $\mathchar"7103_c^+\pi^{-}$ mass and (b) $\cos{\theta_{\mathchar"7103_c^+\pi^{-}}}$ for the full data sample with the results of the fits including the $\mathchar"7106_c(2455)^0$ resonance  with three possible quantum number assignments. The fluctuations for the curves are due to the limited size of the simulation samples. The distributions are not corrected for efficiency.
    }
    \label{fig:valid_2455}
\end{figure*}

An amplitude fit is performed to data in the $\Sigmares_c(2455)^0$ region, including contributions of the $\Sigmares_c(2455)^0$ resonance and the combinatorial background. The spins $1/2$, $3/2$ and $5/2$ are tested for the $\Sigmares_c(2455)^0$ baryon. The best fit is obtained with $J=1/2$, with overwhelming significance compared with the other two hypotheses, and the corresponding result is presented in Fig.~\ref{fig:valid_2455}.
This spin assignment is consistent with both the predictions of the quark model and the results reported by the \babar collaboration~\cite{BaBar:2008get}. Note that, for a single resonance, the parity cannot be measured.

\subsection*{Fit fractions using the baseline models}
Fit fractions of the contributions measured in the Dalitz plot region of \mbox{$m^2_{\Lc\pim}>6.25\gevgevcccc$} and $m^2_{\Lc\antiproton}>16\gevgevcccc$ are shown in Table~\ref{tab:fitfraction}.

\begin{table}[!htbp]
\centering
\renewcommand{\arraystretch}{1.2}
\caption{Fit fraction $\mathcal{F}$ (in \%) of each contribution for the two solution groups. 
}
\label{tab:fitfraction}
\begin{tabular}{lcc}
\hline
Decay channel & Group A &  Group B\\
\hline
$\decay{\Bm}{\Sigmares_c(2520)^0\antiproton}$&$\phantom{0}3.4\pm 0.3\pm0.8$ &$\phantom{0}3.8\pm0.4\pm0.8\phantom{0}$ \\
$\decay{\Bm}{\Sigmares_c(2800)^0\antiproton}$&$\phantom{0}3.4\pm0.7\pm1.6$ & $31.4\pm2.4\pm14.6$ \\
$\decay{\Bm}{\Sigmares_c(2900)^0\antiproton}$&$59.7\pm2.6\pm9.0$ & $17.9\pm1.4\pm5.8\phantom{0}$  \\
$\decay{\Bm}{\Sigmares_c(3200)^0\antiproton}$&$\phantom{0}6.1\pm1.0\pm1.7$ & $\phantom{0}4.7\pm0.7\pm1.8\phantom{0}$  \\
$\decay{\Bm}{(\Lc\pim)_{\text{NR S-wave}}\antiproton}$& $37.7\pm2.4\pm8.2$ & $38.2\pm1.9\pm8.1\phantom{0}$ \\
$\decay{\Bm}{\Deltaresbar(1232)^{--}\Lc}$               &$13.0\pm 1.0\pm3.8$ &$15.1\pm1.2\pm3.5\phantom{0}$ \\
$\decay{\Bm}{\Deltaresbar(1600)^{--}\Lc}$               &$10.3\pm 1.4\pm 5.6$ &$12.5\pm2.4\pm9.0\phantom{0}$ \\
$\decay{\Bm}{\Deltaresbar(1700)^{--}\Lc}$               &$\phantom{0}5.9\pm 1.1\pm1.7$ &$\phantom{0}3.4\pm0.9\pm2.3\phantom{0}$ \\
\hline
\end{tabular}
\end{table}

\subsection*{Systematic uncertainties}
Tables~\ref{tab:syserrA} and~\ref{tab:syserrB} show the systematic uncertainties for the mass, width and relative fit fraction measurements for solutions groups A and B, respectively.

\begin{table}[!htbp]
\centering
\caption{Systematic uncertainties on the mass, width and relative fit fraction for solution \mbox{group A}, for each $\Sigmares_c^0$ state. 
}
\label{tab:syserrA}
\resizebox{1.0\columnwidth}{!}{
\begin{tabular}{l|ccc|ccc|ccc}
\hline
\multirow{3}{*}{Source} & \multicolumn{3}{c|}{$\Sigmares_c(2800)^0$} & \multicolumn{3}{c|}{$\Sigmares_c(2900)^0$} & \multicolumn{3}{c}{$\Sigmares_c(3200)^0$} \\
\cline{2-4} \cline{5-7} \cline{8-10}
& Mass & Width & $\mathcal{R}$ & Mass & Width & $\mathcal{R}$ & Mass & Width & $\mathcal{R}$ \\
& $[\!\mevcc]$ & $[\!\mev]$ & ($ 10^{-3}$) 
& $[\!\mevcc]$ & $[\!\mev]$ & ($ 10^{-3}$) 
& $[\!\mevcc]$ & $[\!\mev]$ & ($ 10^{-3}$)  \\

\hline
Mass fit      & 0.3 & 3.2 & 16 & 1.2 & \phz5.6 & \phantom{0}36  & \phz2.2 & \phz2.1 & \phantom{0}9\\
Efficiency           & 1.2 & 2.8 & 14  & 3.0 &  \phz6.0 & \phantom{0}80  &  \phz4.8 &  16.8 & 33 \\
$\Lc$ decay    & 0.1 & 0.7 & \phantom{0}1  & 0.6 & \phz1.3 & \phantom{0}10  & \phz0.4 & \phz1.9 & \phantom{0}4 \\
Fixed parameters  & 0.6 & 2.4 & 14 & 4.1 & 11.4 & 102 & \phz6.6 & \phz9.4 & 25\\
Alternative model    & 0.3 & 2.2 & \phantom{0}8  & 0.8 & \phz3.4 & \phantom{0}18  & \phz1.4 & \phz3.2 & \phantom{0}5\\
Nonresonant model         & 0.7 & 0.6 & 10 & 2.9 & \phz6.2 & \phantom{0}33  & \phz1.8 & \phz4.1 & 14 \\
Variation of $J^P$  & 1.2 & 5.9 & 36 & 6.3 & 16.1 & 250 & 11.4 & 17.6 & 30 \\
\hline
Total                & 2\phantom{.0}  & 8\phantom{.0}  & 46   & 9\phantom{.0} & 23\phantom{.0} & 286  & 14\phantom{.0} & 27\phantom{.0}  & 54 \\
\hline
\end{tabular}
}
\end{table}

\begin{table}[!htbp]
\centering
\caption{Systematic uncertainties on the mass, width and relative fit fraction for solution \mbox{group B}, for each $\Sigmares_c^0$ state. 
}
\label{tab:syserrB}
\resizebox{1.0\columnwidth}{!}{
\begin{tabular}{l|ccc|ccc|ccc}
\hline
\multirow{3}{*}{Source} & \multicolumn{3}{c|}{$\Sigmares_c(2800)^0$} & \multicolumn{3}{c|}{$\Sigmares_c(2900)^0$} & \multicolumn{3}{c}{$\Sigmares_c(3200)^0$} \\
\cline{2-4} \cline{5-7} \cline{8-10}
& Mass & Width & $\mathcal{R}$ & Mass & Width & $\mathcal{R}$ & Mass & Width & $\mathcal{R}$ \\
& $[\!\mevcc]$ & $[\!\mev]$ & ($ 10^{-3}$) 
& $[\!\mevcc]$ & $[\!\mev]$ & ($ 10^{-3}$) 
& $[\!\mevcc]$ & $[\!\mev]$ & ($ 10^{-3}$) \\

\hline
Mass fit      & 1.3 & \phz3.4 & \phantom{0}56  & 1.9 & \phz2.9 & \phantom{0}11  & 2.1 & \phz3.2 & \phantom{0}1   \\
Efficiency           & 3.6 & \phz7.3 & 180 & 3.2 & \phz8.5 & \phantom{0}86  & 4.2 & \phz9.4 & 32   \\
$\Lc$ decay   & 0.1 & \phz3.7 & 108 & 1.9 & \phz3.6 & \phantom{0}39  & 0.2 & \phz0.8 & \phantom{0}1   \\
Fixed parameters     & 2.8 & 11.5 & 263 & 5.5 & 13.8 & 156 & 6.2 & 25.1 & 33  \\
Alternative model    & 1.9 & \phz5.3 & \phantom{0}36  & 3.0 & \phz4.6 & \phantom{0}\phantom{0}4   & 3.2 & \phz4.9 & \phantom{0}0   \\
Nonresonant model         & 1.7 & \phz6.3 & \phantom{0}74  & 2.4 & \phz2.8 & \phantom{0}14  & 2.7 & \phz0.3 & \phantom{0}3   \\
Variation of $J^P$ & 1.5 & 12.5 & 230 & 1.9 & 15.2 & \phantom{0}60  & 6.1 & 17.1 & 25  \\
\hline
Total                & 6\phantom{0.} & 21\phantom{0.} & 420 & 8\phantom{0.} & 23\phantom{0.} & 193 & 11\phantom{0.} & 32\phantom{0.} & 52 \\
\hline
\end{tabular}
}
\end{table}



\clearpage
\addcontentsline{toc}{section}{References}
\bibliographystyle{LHCb/LHCb}
\bibliography{main,LHCb/standard,LHCb/LHCb-PAPER,LHCb/LHCb-CONF,LHCb/LHCb-DP,LHCb/LHCb-TDR}

\ifx\mcitethebibliography\mciteundefinedmacro
\PackageError{LHCb.bst}{mciteplus.sty has not been loaded}
{This bibstyle requires the use of the mciteplus package.}\fi
\providecommand{\href}[2]{#2}
\begin{mcitethebibliography}{10}
\mciteSetBstSublistMode{n}
\mciteSetBstMaxWidthForm{subitem}{\alph{mcitesubitemcount})}
\mciteSetBstSublistLabelBeginEnd{\mcitemaxwidthsubitemform\space}
{\relax}{\relax}

\bibitem{quarkmodel1}
M.~Gell-Mann, \ifthenelse{\boolean{articletitles}}{\emph{{A schematic model of baryons and mesons}}, }{}\href{https://doi.org/10.1016/S0031-9163(64)92001-3}{Phys.\ Lett.\  \textbf{8} (1964) 214}\relax
\mciteBstWouldAddEndPuncttrue
\mciteSetBstMidEndSepPunct{\mcitedefaultmidpunct}
{\mcitedefaultendpunct}{\mcitedefaultseppunct}\relax
\EndOfBibitem
\bibitem{quarkmodel2}
G.~Zweig, \ifthenelse{\boolean{articletitles}}{\emph{{An SU(3) model for strong interaction symmetry and its breaking. Version 1}}, }{}doi:~\href{https://doi.org/10.17181/CERN-TH-401}{10.17181/CERN-TH-401}\relax
\mciteBstWouldAddEndPuncttrue
\mciteSetBstMidEndSepPunct{\mcitedefaultmidpunct}
{\mcitedefaultendpunct}{\mcitedefaultseppunct}\relax
\EndOfBibitem
\bibitem{Petermann:1965qlk}
A.~Petermann, \ifthenelse{\boolean{articletitles}}{\emph{Propriétés de l'étrangeté et une formule de masse pour les mésons vectoriels}, }{}\href{https://doi.org/10.1016/0029-5582(65)90348-2}{Nucl.\ Phys.\  \textbf{63} (1965) 349}\relax
\mciteBstWouldAddEndPuncttrue
\mciteSetBstMidEndSepPunct{\mcitedefaultmidpunct}
{\mcitedefaultendpunct}{\mcitedefaultseppunct}\relax
\EndOfBibitem
\bibitem{Wang:2024jyk}
E.~Wang {\em et~al.}, \ifthenelse{\boolean{articletitles}}{\emph{{Review of the low-lying excited baryons $\mathit{\Sigma}^*(1/2^-)$}}, }{}\href{https://doi.org/10.1088/0256-307X/41/10/101401}{Chin.\ Phys.\ Lett.\  \textbf{41} (2024) 101401}, \href{http://arxiv.org/abs/2406.07839}{{\normalfont\ttfamily arXiv:2406.07839}}\relax
\mciteBstWouldAddEndPuncttrue
\mciteSetBstMidEndSepPunct{\mcitedefaultmidpunct}
{\mcitedefaultendpunct}{\mcitedefaultseppunct}\relax
\EndOfBibitem
\bibitem{Chen:2022asf}
H.-X. Chen {\em et~al.}, \ifthenelse{\boolean{articletitles}}{\emph{{An updated review of the new hadron states}}, }{}\href{https://doi.org/10.1088/1361-6633/aca3b6}{Rept.\ Prog.\ Phys.\  \textbf{86} (2023) 026201}, \href{http://arxiv.org/abs/2204.02649}{{\normalfont\ttfamily arXiv:2204.02649}}\relax
\mciteBstWouldAddEndPuncttrue
\mciteSetBstMidEndSepPunct{\mcitedefaultmidpunct}
{\mcitedefaultendpunct}{\mcitedefaultseppunct}\relax
\EndOfBibitem
\bibitem{Luo:2025sns}
X.~Luo {\em et~al.}, \ifthenelse{\boolean{articletitles}}{\emph{{A short review on QCD sum rule studies of P-wave single heavy baryons}}, }{}\href{http://arxiv.org/abs/2510.13013}{{\normalfont\ttfamily arXiv:2510.13013}}\relax
\mciteBstWouldAddEndPuncttrue
\mciteSetBstMidEndSepPunct{\mcitedefaultmidpunct}
{\mcitedefaultendpunct}{\mcitedefaultseppunct}\relax
\EndOfBibitem
\bibitem{PDG2024}
Particle Data Group, S.~Navas {\em et~al.}, \ifthenelse{\boolean{articletitles}}{\emph{{\href{http://pdg.lbl.gov/}{Review of particle physics}}}, }{}\href{https://doi.org/10.1103/PhysRevD.110.030001}{Phys.\ Rev.\  \textbf{D110} (2024) 030001}\relax
\mciteBstWouldAddEndPuncttrue
\mciteSetBstMidEndSepPunct{\mcitedefaultmidpunct}
{\mcitedefaultendpunct}{\mcitedefaultseppunct}\relax
\EndOfBibitem
\bibitem{CLEO:1996czm}
CLEO collaboration, G.~Brandenburg {\em et~al.}, \ifthenelse{\boolean{articletitles}}{\emph{{Observation of two excited charmed baryons decaying into ${\ensuremath{\mathit{\Lambda}}}_{\mathit{c}}^{+}{\ensuremath{\mathit{\pi}}}^{\ifmmode\pm\else\textpm\fi{}}$}}, }{}\href{https://doi.org/10.1103/PhysRevLett.78.2304}{Phys.\ Rev.\ Lett.\  \textbf{78} (1997) 2304}\relax
\mciteBstWouldAddEndPuncttrue
\mciteSetBstMidEndSepPunct{\mcitedefaultmidpunct}
{\mcitedefaultendpunct}{\mcitedefaultseppunct}\relax
\EndOfBibitem
\bibitem{CLEO:2000abd}
CLEO collaboration, R.~Ammar {\em et~al.}, \ifthenelse{\boolean{articletitles}}{\emph{{First observation of the ${\ensuremath{\mathit{\Sigma}}}_{\mathit{c}}^{*+}$ baryon and a new measurement of the ${\ensuremath{\mathit{\Sigma}}}_{\mathit{c}}^{*+}$ mass}}, }{}\href{https://doi.org/10.1103/PhysRevLett.86.1167}{Phys.\ Rev.\ Lett.\  \textbf{86} (2001) 1167}, \href{http://arxiv.org/abs/hep-ex/0007041}{{\normalfont\ttfamily arXiv:hep-ex/0007041}}\relax
\mciteBstWouldAddEndPuncttrue
\mciteSetBstMidEndSepPunct{\mcitedefaultmidpunct}
{\mcitedefaultendpunct}{\mcitedefaultseppunct}\relax
\EndOfBibitem
\bibitem{Belle:2004zjl}
Belle collaboration, R.~Mizuk {\em et~al.}, \ifthenelse{\boolean{articletitles}}{\emph{{Observation of an isotriplet of excited charmed baryons decaying to $\Lc \pi$}}, }{}\href{https://doi.org/10.1103/PhysRevLett.94.122002}{Phys.\ Rev.\ Lett.\  \textbf{94} (2005) 122002}, \href{http://arxiv.org/abs/hep-ex/0412069}{{\normalfont\ttfamily arXiv:hep-ex/0412069}}\relax
\mciteBstWouldAddEndPuncttrue
\mciteSetBstMidEndSepPunct{\mcitedefaultmidpunct}
{\mcitedefaultendpunct}{\mcitedefaultseppunct}\relax
\EndOfBibitem
\bibitem{Chen:2016spr}
H.-X. Chen {\em et~al.}, \ifthenelse{\boolean{articletitles}}{\emph{{A review of the open charm and open bottom systems}}, }{}\href{https://doi.org/10.1088/1361-6633/aa6420}{Rept.\ Prog.\ Phys.\  \textbf{80} (2017) 076201}, \href{http://arxiv.org/abs/1609.08928}{{\normalfont\ttfamily arXiv:1609.08928}}\relax
\mciteBstWouldAddEndPuncttrue
\mciteSetBstMidEndSepPunct{\mcitedefaultmidpunct}
{\mcitedefaultendpunct}{\mcitedefaultseppunct}\relax
\EndOfBibitem
\bibitem{Cheng:2021qpd}
H.-Y. Cheng, \ifthenelse{\boolean{articletitles}}{\emph{{Charmed baryon physics circa 2021}}, }{}\href{https://doi.org/10.1016/j.cjph.2022.06.021}{Chin.\ J.\ Phys.\  \textbf{78} (2022) 324}, \href{http://arxiv.org/abs/2109.01216}{{\normalfont\ttfamily arXiv:2109.01216}}\relax
\mciteBstWouldAddEndPuncttrue
\mciteSetBstMidEndSepPunct{\mcitedefaultmidpunct}
{\mcitedefaultendpunct}{\mcitedefaultseppunct}\relax
\EndOfBibitem
\bibitem{WangWangKaiLei:2021kdd}
K.-L. Wang and X.-H. Zhong, \ifthenelse{\boolean{articletitles}}{\emph{{Toward discovering low-lying $P$-wave excited \Sigmac baryon states}}, }{}\href{https://doi.org/10.1088/1674-1137/ac3123}{Chin.\ Phys.\  \textbf{C46} (2022) 023103}, \href{http://arxiv.org/abs/2110.12443}{{\normalfont\ttfamily arXiv:2110.12443}}\relax
\mciteBstWouldAddEndPuncttrue
\mciteSetBstMidEndSepPunct{\mcitedefaultmidpunct}
{\mcitedefaultendpunct}{\mcitedefaultseppunct}\relax
\EndOfBibitem
\bibitem{BaBar:2008get}
BaBar collaboration, B.~Aubert {\em et~al.}, \ifthenelse{\boolean{articletitles}}{\emph{{Measurements of $\BF(\Bzb\to\Lc\antiproton)$ and $\BF(\Bm\to\Lc\antiproton\pim)$ and studies of $\Lc\pim$ resonances}}, }{}\href{https://doi.org/10.1103/PhysRevD.78.112003}{Phys.\ Rev.\  \textbf{D78} (2008) 112003}, \href{http://arxiv.org/abs/0807.4974}{{\normalfont\ttfamily arXiv:0807.4974}}\relax
\mciteBstWouldAddEndPuncttrue
\mciteSetBstMidEndSepPunct{\mcitedefaultmidpunct}
{\mcitedefaultendpunct}{\mcitedefaultseppunct}\relax
\EndOfBibitem
\bibitem{Li:2024zze}
Z.-Y. Li, G.-L. Yu, Z.-G. Wang, and J.-Z. Gu, \ifthenelse{\boolean{articletitles}}{\emph{{Heavy-quark dominance and fine structure of excited heavy baryons $\mathit{\Sigma} _{Q}$, $\mathit{\Xi} '_{Q}$ and $\mathit{\Omega} _{Q}$}}, }{}\href{https://doi.org/10.1140/epjc/s10052-024-13706-5}{Eur.\ Phys.\ J.\  \textbf{C84} (2024) 1310}, \href{http://arxiv.org/abs/2405.16162}{{\normalfont\ttfamily arXiv:2405.16162}}\relax
\mciteBstWouldAddEndPuncttrue
\mciteSetBstMidEndSepPunct{\mcitedefaultmidpunct}
{\mcitedefaultendpunct}{\mcitedefaultseppunct}\relax
\EndOfBibitem
\bibitem{Garcilazo:2007eh}
H.~Garcilazo, J.~Vijande, and A.~Valcarce, \ifthenelse{\boolean{articletitles}}{\emph{{Faddeev study of heavy baryon spectroscopy}}, }{}\href{https://doi.org/10.1088/0954-3899/34/5/014}{J.\ Phys.\  \textbf{G34} (2007) 961}, \href{http://arxiv.org/abs/hep-ph/0703257}{{\normalfont\ttfamily arXiv:hep-ph/0703257}}\relax
\mciteBstWouldAddEndPuncttrue
\mciteSetBstMidEndSepPunct{\mcitedefaultmidpunct}
{\mcitedefaultendpunct}{\mcitedefaultseppunct}\relax
\EndOfBibitem
\bibitem{Valcarce:2008dr}
A.~Valcarce, H.~Garcilazo, and J.~Vijande, \ifthenelse{\boolean{articletitles}}{\emph{{Towards an understanding of heavy baryon spectroscopy}}, }{}\href{https://doi.org/10.1140/epja/i2008-10616-4}{Eur.\ Phys.\ J.\  \textbf{A37} (2008) 217}, \href{http://arxiv.org/abs/0807.2973}{{\normalfont\ttfamily arXiv:0807.2973}}\relax
\mciteBstWouldAddEndPuncttrue
\mciteSetBstMidEndSepPunct{\mcitedefaultmidpunct}
{\mcitedefaultendpunct}{\mcitedefaultseppunct}\relax
\EndOfBibitem
\bibitem{Jakhad:2023mni}
P.~Jakhad, J.~Oudichhya, K.~Gandhi, and A.~K. Rai, \ifthenelse{\boolean{articletitles}}{\emph{{Identification of newly observed singly charmed baryons using the relativistic flux tube model}}, }{}\href{https://doi.org/10.1103/PhysRevD.108.014011}{Phys.\ Rev.\  \textbf{D108} (2023) 014011}, \href{http://arxiv.org/abs/2306.06349}{{\normalfont\ttfamily arXiv:2306.06349}}\relax
\mciteBstWouldAddEndPuncttrue
\mciteSetBstMidEndSepPunct{\mcitedefaultmidpunct}
{\mcitedefaultendpunct}{\mcitedefaultseppunct}\relax
\EndOfBibitem
\bibitem{Luo:2023sne}
S.-Q. Luo and X.~Liu, \ifthenelse{\boolean{articletitles}}{\emph{{Investigating the spectroscopy behavior of undetected $1$F-wave charmed baryons}}, }{}\href{https://doi.org/10.1103/PhysRevD.108.034002}{Phys.\ Rev.\  \textbf{D108} (2023) 034002}, \href{http://arxiv.org/abs/2306.04588}{{\normalfont\ttfamily arXiv:2306.04588}}\relax
\mciteBstWouldAddEndPuncttrue
\mciteSetBstMidEndSepPunct{\mcitedefaultmidpunct}
{\mcitedefaultendpunct}{\mcitedefaultseppunct}\relax
\EndOfBibitem
\bibitem{Yang:2021lce}
H.-M. Yang and H.-X. Chen, \ifthenelse{\boolean{articletitles}}{\emph{{$P$-wave charmed baryons of the $SU(3)$ flavor $6_F$}}, }{}\href{https://doi.org/10.1103/PhysRevD.104.034037}{Phys.\ Rev.\  \textbf{D104} (2021) 034037}, \href{http://arxiv.org/abs/2106.15488}{{\normalfont\ttfamily arXiv:2106.15488}}\relax
\mciteBstWouldAddEndPuncttrue
\mciteSetBstMidEndSepPunct{\mcitedefaultmidpunct}
{\mcitedefaultendpunct}{\mcitedefaultseppunct}\relax
\EndOfBibitem
\bibitem{Chen:2016iyi}
B.~Chen, K.-W. Wei, X.~Liu, and T.~Matsuki, \ifthenelse{\boolean{articletitles}}{\emph{{Low-lying charmed and charmed-strange baryon states}}, }{}\href{https://doi.org/10.1140/epjc/s10052-017-4708-x}{Eur.\ Phys.\ J.\  \textbf{C77} (2017) 154}, \href{http://arxiv.org/abs/1609.07967}{{\normalfont\ttfamily arXiv:1609.07967}}\relax
\mciteBstWouldAddEndPuncttrue
\mciteSetBstMidEndSepPunct{\mcitedefaultmidpunct}
{\mcitedefaultendpunct}{\mcitedefaultseppunct}\relax
\EndOfBibitem
\bibitem{Ebert:2011kk}
D.~Ebert, R.~N. Faustov, and V.~O. Galkin, \ifthenelse{\boolean{articletitles}}{\emph{{Spectroscopy and Regge trajectories of heavy baryons in the relativistic quark-diquark picture}}, }{}\href{https://doi.org/10.1103/PhysRevD.84.014025}{Phys.\ Rev.\  \textbf{D84} (2011) 014025}, \href{http://arxiv.org/abs/1105.0583}{{\normalfont\ttfamily arXiv:1105.0583}}\relax
\mciteBstWouldAddEndPuncttrue
\mciteSetBstMidEndSepPunct{\mcitedefaultmidpunct}
{\mcitedefaultendpunct}{\mcitedefaultseppunct}\relax
\EndOfBibitem
\bibitem{Shah:2016mig}
Z.~Shah, K.~Thakkar, A.~Kumar~Rai, and P.~C. Vinodkumar, \ifthenelse{\boolean{articletitles}}{\emph{{Excited state mass spectra of singly charmed baryons}}, }{}\href{https://doi.org/10.1140/epja/i2016-16313-9}{Eur.\ Phys.\ J.\  \textbf{A52} (2016) 313}, \href{http://arxiv.org/abs/1602.06384}{{\normalfont\ttfamily arXiv:1602.06384}}\relax
\mciteBstWouldAddEndPuncttrue
\mciteSetBstMidEndSepPunct{\mcitedefaultmidpunct}
{\mcitedefaultendpunct}{\mcitedefaultseppunct}\relax
\EndOfBibitem
\bibitem{LHCb-DP-2008-001}
LHCb collaboration, A.~A. Alves~Jr.\ {\em et~al.}, \ifthenelse{\boolean{articletitles}}{\emph{{The \lhcb detector at the LHC}}, }{}\href{https://doi.org/10.1088/1748-0221/3/08/S08005}{JINST \textbf{3} (2008) S08005}\relax
\mciteBstWouldAddEndPuncttrue
\mciteSetBstMidEndSepPunct{\mcitedefaultmidpunct}
{\mcitedefaultendpunct}{\mcitedefaultseppunct}\relax
\EndOfBibitem
\bibitem{LHCb-DP-2014-002}
LHCb collaboration, R.~Aaij {\em et~al.}, \ifthenelse{\boolean{articletitles}}{\emph{{LHCb detector performance}}, }{}\href{https://doi.org/10.1142/S0217751X15300227}{Int.\ J.\ Mod.\ Phys.\  \textbf{A30} (2015) 1530022}, \href{http://arxiv.org/abs/1412.6352}{{\normalfont\ttfamily arXiv:1412.6352}}\relax
\mciteBstWouldAddEndPuncttrue
\mciteSetBstMidEndSepPunct{\mcitedefaultmidpunct}
{\mcitedefaultendpunct}{\mcitedefaultseppunct}\relax
\EndOfBibitem
\bibitem{Sjostrand:2007gs}
T.~Sj\"{o}strand, S.~Mrenna, and P.~Skands, \ifthenelse{\boolean{articletitles}}{\emph{{A brief introduction to PYTHIA 8.1}}, }{}\href{https://doi.org/10.1016/j.cpc.2008.01.036}{Comput.\ Phys.\ Commun.\  \textbf{178} (2008) 852}, \href{http://arxiv.org/abs/0710.3820}{{\normalfont\ttfamily arXiv:0710.3820}}\relax
\mciteBstWouldAddEndPuncttrue
\mciteSetBstMidEndSepPunct{\mcitedefaultmidpunct}
{\mcitedefaultendpunct}{\mcitedefaultseppunct}\relax
\EndOfBibitem
\bibitem{LHCb-PROC-2010-056}
I.~Belyaev {\em et~al.}, \ifthenelse{\boolean{articletitles}}{\emph{{Handling of the generation of primary events in Gauss, the LHCb simulation framework}}, }{}\href{https://doi.org/10.1088/1742-6596/331/3/032047}{J.\ Phys.\ Conf.\ Ser.\  \textbf{331} (2011) 032047}\relax
\mciteBstWouldAddEndPuncttrue
\mciteSetBstMidEndSepPunct{\mcitedefaultmidpunct}
{\mcitedefaultendpunct}{\mcitedefaultseppunct}\relax
\EndOfBibitem
\bibitem{Lange:2001uf}
D.~J. Lange, \ifthenelse{\boolean{articletitles}}{\emph{{The EvtGen particle decay simulation package}}, }{}\href{https://doi.org/10.1016/S0168-9002(01)00089-4}{Nucl.\ Instrum.\ Meth.\  \textbf{A462} (2001) 152}\relax
\mciteBstWouldAddEndPuncttrue
\mciteSetBstMidEndSepPunct{\mcitedefaultmidpunct}
{\mcitedefaultendpunct}{\mcitedefaultseppunct}\relax
\EndOfBibitem
\bibitem{davidson2015photos}
N.~Davidson, T.~Przedzinski, and Z.~Was, \ifthenelse{\boolean{articletitles}}{\emph{{PHOTOS interface in C++: Technical and physics documentation}}, }{}\href{https://doi.org/https://doi.org/10.1016/j.cpc.2015.09.013}{Comput.\ Phys.\ Commun.\  \textbf{199} (2016) 86}, \href{http://arxiv.org/abs/1011.0937}{{\normalfont\ttfamily arXiv:1011.0937}}\relax
\mciteBstWouldAddEndPuncttrue
\mciteSetBstMidEndSepPunct{\mcitedefaultmidpunct}
{\mcitedefaultendpunct}{\mcitedefaultseppunct}\relax
\EndOfBibitem
\bibitem{Allison:2006ve}
Geant4 collaboration, J.~Allison {\em et~al.}, \ifthenelse{\boolean{articletitles}}{\emph{{Geant4 developments and applications}}, }{}\href{https://doi.org/10.1109/TNS.2006.869826}{IEEE Trans.\ Nucl.\ Sci.\  \textbf{53} (2006) 270}\relax
\mciteBstWouldAddEndPuncttrue
\mciteSetBstMidEndSepPunct{\mcitedefaultmidpunct}
{\mcitedefaultendpunct}{\mcitedefaultseppunct}\relax
\EndOfBibitem
\bibitem{Agostinelli:2002hh}
Geant4 collaboration, S.~Agostinelli {\em et~al.}, \ifthenelse{\boolean{articletitles}}{\emph{{Geant4: A simulation toolkit}}, }{}\href{https://doi.org/10.1016/S0168-9002(03)01368-8}{Nucl.\ Instrum.\ Meth.\  \textbf{A506} (2003) 250}\relax
\mciteBstWouldAddEndPuncttrue
\mciteSetBstMidEndSepPunct{\mcitedefaultmidpunct}
{\mcitedefaultendpunct}{\mcitedefaultseppunct}\relax
\EndOfBibitem
\bibitem{LHCb-PROC-2011-006}
M.~Clemencic {\em et~al.}, \ifthenelse{\boolean{articletitles}}{\emph{{The \lhcb simulation application, Gauss: Design, evolution and experience}}, }{}\href{https://doi.org/10.1088/1742-6596/331/3/032023}{J.\ Phys.\ Conf.\ Ser.\  \textbf{331} (2011) 032023}\relax
\mciteBstWouldAddEndPuncttrue
\mciteSetBstMidEndSepPunct{\mcitedefaultmidpunct}
{\mcitedefaultendpunct}{\mcitedefaultseppunct}\relax
\EndOfBibitem
\bibitem{LHCb-PUB-2016-021}
L.~Anderlini {\em et~al.}, \ifthenelse{\boolean{articletitles}}{\emph{{The PIDCalib package}}, }{} \href{http://cdsweb.cern.ch/search?p=LHCb-PUB-2016-021&f=reportnumber&action_search=Search&c=LHCb+Notes} {LHCb-PUB-2016-021}, 2016\relax
\mciteBstWouldAddEndPuncttrue
\mciteSetBstMidEndSepPunct{\mcitedefaultmidpunct}
{\mcitedefaultendpunct}{\mcitedefaultseppunct}\relax
\EndOfBibitem
\bibitem{LHCb-DP-2013-002}
LHCb collaboration, R.~Aaij {\em et~al.}, \ifthenelse{\boolean{articletitles}}{\emph{{Measurement of the track reconstruction efficiency at LHCb}}, }{}\href{https://doi.org/10.1088/1748-0221/10/02/P02007}{JINST \textbf{10} (2015) P02007}, \href{http://arxiv.org/abs/1408.1251}{{\normalfont\ttfamily arXiv:1408.1251}}\relax
\mciteBstWouldAddEndPuncttrue
\mciteSetBstMidEndSepPunct{\mcitedefaultmidpunct}
{\mcitedefaultendpunct}{\mcitedefaultseppunct}\relax
\EndOfBibitem
\bibitem{LHCb-DP-2020-001}
C.~Abell\'an~Beteta {\em et~al.}, \ifthenelse{\boolean{articletitles}}{\emph{{Calibration and performance of the LHCb calorimeters in Run 1 and 2 at the LHC}}, }{}\href{http://arxiv.org/abs/2008.11556}{{\normalfont\ttfamily arXiv:2008.11556}}\relax
\mciteBstWouldAddEndPuncttrue
\mciteSetBstMidEndSepPunct{\mcitedefaultmidpunct}
{\mcitedefaultendpunct}{\mcitedefaultseppunct}\relax
\EndOfBibitem
\bibitem{LHCb-DP-2018-001}
R.~Aaij {\em et~al.}, \ifthenelse{\boolean{articletitles}}{\emph{{Selection and processing of calibration samples to measure the particle identification performance of the LHCb experiment in Run 2}}, }{}\href{https://doi.org/10.1140/epjti/s40485-019-0050-z}{Eur.\ Phys.\ J.\ Tech.\ Instr.\  \textbf{6} (2019) 1}, \href{http://arxiv.org/abs/1803.00824}{{\normalfont\ttfamily arXiv:1803.00824}}\relax
\mciteBstWouldAddEndPuncttrue
\mciteSetBstMidEndSepPunct{\mcitedefaultmidpunct}
{\mcitedefaultendpunct}{\mcitedefaultseppunct}\relax
\EndOfBibitem
\bibitem{LHCb-DP-2012-004}
R.~Aaij {\em et~al.}, \ifthenelse{\boolean{articletitles}}{\emph{{The \lhcb trigger and its performance in 2011}}, }{}\href{https://doi.org/10.1088/1748-0221/8/04/P04022}{JINST \textbf{8} (2013) P04022}, \href{http://arxiv.org/abs/1211.3055}{{\normalfont\ttfamily arXiv:1211.3055}}\relax
\mciteBstWouldAddEndPuncttrue
\mciteSetBstMidEndSepPunct{\mcitedefaultmidpunct}
{\mcitedefaultendpunct}{\mcitedefaultseppunct}\relax
\EndOfBibitem
\bibitem{LHCb-DP-2019-001}
R.~Aaij {\em et~al.}, \ifthenelse{\boolean{articletitles}}{\emph{{Design and performance of the LHCb trigger and full real-time reconstruction in Run 2 of the LHC}}, }{}\href{https://doi.org/10.1088/1748-0221/14/04/P04013}{JINST \textbf{14} (2019) P04013}, \href{http://arxiv.org/abs/1812.10790}{{\normalfont\ttfamily arXiv:1812.10790}}\relax
\mciteBstWouldAddEndPuncttrue
\mciteSetBstMidEndSepPunct{\mcitedefaultmidpunct}
{\mcitedefaultendpunct}{\mcitedefaultseppunct}\relax
\EndOfBibitem
\bibitem{Hulsbergen:2005pu}
W.~D. Hulsbergen, \ifthenelse{\boolean{articletitles}}{\emph{{Decay chain fitting with a Kalman filter}}, }{}\href{https://doi.org/10.1016/j.nima.2005.06.078}{Nucl.\ Instrum.\ Meth.\  \textbf{A552} (2005) 566}, \href{http://arxiv.org/abs/physics/0503191}{{\normalfont\ttfamily arXiv:physics/0503191}}\relax
\mciteBstWouldAddEndPuncttrue
\mciteSetBstMidEndSepPunct{\mcitedefaultmidpunct}
{\mcitedefaultendpunct}{\mcitedefaultseppunct}\relax
\EndOfBibitem
\bibitem{Breiman}
L.~Breiman, J.~H. Friedman, R.~A. Olshen, and C.~J. Stone, {\em Classification and regression trees}, Wadsworth international group, Belmont, California, USA, 1984\relax
\mciteBstWouldAddEndPuncttrue
\mciteSetBstMidEndSepPunct{\mcitedefaultmidpunct}
{\mcitedefaultendpunct}{\mcitedefaultseppunct}\relax
\EndOfBibitem
\bibitem{AdaBoost}
Y.~Freund and R.~E. Schapire, \ifthenelse{\boolean{articletitles}}{\emph{A decision-theoretic generalization of on-line learning and an application to boosting}, }{}\href{https://doi.org/10.1006/jcss.1997.1504}{J.\ Comput.\ Syst.\ Sci.\  \textbf{55} (1997) 119}\relax
\mciteBstWouldAddEndPuncttrue
\mciteSetBstMidEndSepPunct{\mcitedefaultmidpunct}
{\mcitedefaultendpunct}{\mcitedefaultseppunct}\relax
\EndOfBibitem
\bibitem{Hocker:2007ht}
H.~Voss, A.~Hoecker, J.~Stelzer, and F.~Tegenfeldt, \ifthenelse{\boolean{articletitles}}{\emph{{TMVA - Toolkit for Multivariate Data Analysis with ROOT}}, }{}\href{https://doi.org/10.22323/1.050.0040}{PoS \textbf{ACAT} (2007) 040}\relax
\mciteBstWouldAddEndPuncttrue
\mciteSetBstMidEndSepPunct{\mcitedefaultmidpunct}
{\mcitedefaultendpunct}{\mcitedefaultseppunct}\relax
\EndOfBibitem
\bibitem{TMVA4}
A.~Hoecker {\em et~al.}, \ifthenelse{\boolean{articletitles}}{\emph{{TMVA 4 --- Toolkit for Multivariate Data Analysis with ROOT. Users Guide.}}, }{}\href{http://arxiv.org/abs/physics/0703039}{{\normalfont\ttfamily arXiv:physics/0703039}}\relax
\mciteBstWouldAddEndPuncttrue
\mciteSetBstMidEndSepPunct{\mcitedefaultmidpunct}
{\mcitedefaultendpunct}{\mcitedefaultseppunct}\relax
\EndOfBibitem
\bibitem{Skwarnicki:1986xj}
T.~Skwarnicki, {\em {A study of the radiative cascade transitions between the Upsilon-prime and Upsilon resonances}}, PhD thesis, Institute of Nuclear Physics, Krakow, 1986, {\href{http://inspirehep.net/record/230779/}{DESY-F31-86-02}}\relax
\mciteBstWouldAddEndPuncttrue
\mciteSetBstMidEndSepPunct{\mcitedefaultmidpunct}
{\mcitedefaultendpunct}{\mcitedefaultseppunct}\relax
\EndOfBibitem
\bibitem{Argus:1990hfq}
ARGUS collaboration, H.~Albrecht {\em et~al.}, \ifthenelse{\boolean{articletitles}}{\emph{{Search for hadronic $b \to u$ decays}}, }{}\href{https://doi.org/10.1016/0370-2693(90)91293-K}{Phys.\ Lett.\  \textbf{B241} (1990) 278}\relax
\mciteBstWouldAddEndPuncttrue
\mciteSetBstMidEndSepPunct{\mcitedefaultmidpunct}
{\mcitedefaultendpunct}{\mcitedefaultseppunct}\relax
\EndOfBibitem
\bibitem{Chua:2001vh}
C.-K. Chua, W.-S. Hou, and S.-Y. Tsai, \ifthenelse{\boolean{articletitles}}{\emph{{Understanding \decay{B}{D^{*-} N\overline N} and its implications}}, }{}\href{https://doi.org/10.1103/PhysRevD.65.034003}{Phys.\ Rev.\  \textbf{D65} (2002) 034003}, \href{http://arxiv.org/abs/hep-ph/0107110}{{\normalfont\ttfamily arXiv:hep-ph/0107110}}\relax
\mciteBstWouldAddEndPuncttrue
\mciteSetBstMidEndSepPunct{\mcitedefaultmidpunct}
{\mcitedefaultendpunct}{\mcitedefaultseppunct}\relax
\EndOfBibitem
\bibitem{Chua:2002wn}
C.-K. Chua, W.-S. Hou, and S.-Y. Tsai, \ifthenelse{\boolean{articletitles}}{\emph{{Charmless three-body baryonic B decays}}, }{}\href{https://doi.org/10.1103/PhysRevD.66.054004}{Phys.\ Rev.\  \textbf{D66} (2002) 054004}, \href{http://arxiv.org/abs/hep-ph/0204185}{{\normalfont\ttfamily arXiv:hep-ph/0204185}}\relax
\mciteBstWouldAddEndPuncttrue
\mciteSetBstMidEndSepPunct{\mcitedefaultmidpunct}
{\mcitedefaultendpunct}{\mcitedefaultseppunct}\relax
\EndOfBibitem
\bibitem{JACOB1959404}
M.~Jacob and G.~C. Wick, \ifthenelse{\boolean{articletitles}}{\emph{{On the general theory of collisions for particles with spin}}, }{}\href{https://doi.org/10.1006/aphy.2000.6022}{Annals Phys.\  \textbf{7} (1959) 404}\relax
\mciteBstWouldAddEndPuncttrue
\mciteSetBstMidEndSepPunct{\mcitedefaultmidpunct}
{\mcitedefaultendpunct}{\mcitedefaultseppunct}\relax
\EndOfBibitem
\bibitem{multiplesolution}
Y.~Gao {\em et~al.}, \ifthenelse{\boolean{articletitles}}{\emph{{A scheme to fix multiple solutions in amplitude analyses}}, }{}\href{https://doi.org/10.1088/1674-1137/ad2674}{Chin.\ Phys.\  \textbf{C48} (2024) 053001}, \href{http://arxiv.org/abs/2302.13862}{{\normalfont\ttfamily arXiv:2302.13862}}\relax
\mciteBstWouldAddEndPuncttrue
\mciteSetBstMidEndSepPunct{\mcitedefaultmidpunct}
{\mcitedefaultendpunct}{\mcitedefaultseppunct}\relax
\EndOfBibitem
\bibitem{Back:2017zqt}
J.~Back {\em et~al.}, \ifthenelse{\boolean{articletitles}}{\emph{{{\sc Laura}$^{++}$: A Dalitz plot fitter}}, }{}\href{https://doi.org/10.1016/j.cpc.2018.04.017}{Comput.\ Phys.\ Commun.\  \textbf{231} (2018) 198}, \href{http://arxiv.org/abs/1711.09854}{{\normalfont\ttfamily arXiv:1711.09854}}\relax
\mciteBstWouldAddEndPuncttrue
\mciteSetBstMidEndSepPunct{\mcitedefaultmidpunct}
{\mcitedefaultendpunct}{\mcitedefaultseppunct}\relax
\EndOfBibitem
\bibitem{LHCb-PAPER-2020-025}
LHCb collaboration, R.~Aaij {\em et~al.}, \ifthenelse{\boolean{articletitles}}{\emph{{Amplitude analysis of the $\Bp \to \Dp \Dm \Kp$ decay}}, }{}\href{https://doi.org/10.1103/PhysRevD.102.112003}{Phys.\ Rev.\  \textbf{D102} (2020) 112003}, \href{http://arxiv.org/abs/2009.00026}{{\normalfont\ttfamily arXiv:2009.00026}}\relax
\mciteBstWouldAddEndPuncttrue
\mciteSetBstMidEndSepPunct{\mcitedefaultmidpunct}
{\mcitedefaultendpunct}{\mcitedefaultseppunct}\relax
\EndOfBibitem
\bibitem{Chung:1995dx}
S.~U. Chung {\em et~al.}, \ifthenelse{\boolean{articletitles}}{\emph{{Partial wave analysis in K matrix formalism}}, }{}\href{https://doi.org/10.1002/andp.19955070504}{Annalen Phys.\  \textbf{4} (1995) 404}\relax
\mciteBstWouldAddEndPuncttrue
\mciteSetBstMidEndSepPunct{\mcitedefaultmidpunct}
{\mcitedefaultendpunct}{\mcitedefaultseppunct}\relax
\EndOfBibitem
\bibitem{Capstick:1986ter}
S.~Capstick and N.~Isgur, \ifthenelse{\boolean{articletitles}}{\emph{{Baryons in a relativized quark model with chromodynamics}}, }{}\href{https://doi.org/10.1103/physrevd.34.2809}{Phys.\ Rev.\  \textbf{D34} (1986) 2809}\relax
\mciteBstWouldAddEndPuncttrue
\mciteSetBstMidEndSepPunct{\mcitedefaultmidpunct}
{\mcitedefaultendpunct}{\mcitedefaultseppunct}\relax
\EndOfBibitem
\bibitem{Ebert:2007nw}
D.~Ebert, R.~N. Faustov, and V.~O. Galkin, \ifthenelse{\boolean{articletitles}}{\emph{{Masses of excited heavy baryons in the relativistic quark model}}, }{}\href{https://doi.org/10.1016/j.physletb.2007.11.037}{Phys.\ Lett.\  \textbf{B659} (2008) 612}, \href{http://arxiv.org/abs/0705.2957}{{\normalfont\ttfamily arXiv:0705.2957}}\relax
\mciteBstWouldAddEndPuncttrue
\mciteSetBstMidEndSepPunct{\mcitedefaultmidpunct}
{\mcitedefaultendpunct}{\mcitedefaultseppunct}\relax
\EndOfBibitem
\end{mcitethebibliography}

\newpage
\centerline
{\large\bf LHCb collaboration}
\begin
{flushleft}
\small
R.~Aaij$^{38}$\lhcborcid{0000-0003-0533-1952},
M.~Abdelfatah$^{69}$,
A.S.W.~Abdelmotteleb$^{57}$\lhcborcid{0000-0001-7905-0542},
C.~Abellan~Beteta$^{51}$\lhcborcid{0009-0009-0869-6798},
F.~Abudin\'en$^{59}$\lhcborcid{0000-0002-6737-3528},
T.~Ackernley$^{61}$\lhcborcid{0000-0002-5951-3498},
A.A.~Adefisoye$^{69}$\lhcborcid{0000-0003-2448-1550},
B.~Adeva$^{47}$\lhcborcid{0000-0001-9756-3712},
M.~Adinolfi$^{55}$\lhcborcid{0000-0002-1326-1264},
P.~Adlarson$^{87}$\lhcborcid{0000-0001-6280-3851}
% ...

{\footnotesize \it

% ...
$^{38}$Nikhef National Institute for Subatomic Physics, Amsterdam, Netherlands\\
% ...
$^{47}$Instituto Galego de F{\'\i}sica de Altas Enerx{\'\i}as (IGFAE), Universidade de Santiago de Compostela, Santiago de Compostela, Spain\\
% ...
$^{51}$Physik-Institut, Universit{\"a}t Z{\"u}rich, Z{\"u}rich, Switzerland\\
% ...
$^{55}$H.H. Wills Physics Laboratory, University of Bristol, Bristol, United Kingdom\\
% ...
$^{57}$Department of Physics, University of Warwick, Coventry, United Kingdom\\
% ...
$^{59}$School of Physics and Astronomy, University of Edinburgh, Edinburgh, United Kingdom\\
$^{60}$School of Physics and Astronomy, University of Glasgow, Glasgow, United Kingdom\\
$^{61}$Oliver Lodge Laboratory, University of Liverpool, Liverpool, United Kingdom\\
% ...
$^{69}$Syracuse University, Syracuse, NY, United States\\
% ...
$^{87}$Department of Physics and Astronomy, Uppsala University, Uppsala, Sweden, associated to $^{60}$\\
% ...
}
\end{flushleft}




\end{document}